\documentclass[
 reprint,
 superscriptaddress,
 amsmath,amssymb,
 aps,
 pra,
]{revtex4-2}

\usepackage{graphicx}
\usepackage{dcolumn}
\usepackage{bm}
\usepackage{hyperref}
\usepackage{siunitx}
\usepackage{xspace}
\usepackage{xcolor}

\newcommand{\ket}[1]{\ensuremath{\lvert #1 \rangle}\xspace}%
\newcommand{\bra}[1]{\ensuremath{\langle #1 \rvert}\xspace}%
\newcommand{\braket}[2]{\ensuremath{\langle #1 \rvert #2\rangle}\xspace}%

\begin{document}

\title{A subwavelength atomic array switched by a single Rydberg atom}

\author{Kritsana~Srakaew}
    \affiliation{Max-Planck-Institut f\"{u}r Quantenoptik, 85748 Garching, Germany}
    \affiliation{Munich Center for Quantum Science and Technology (MCQST), 80799 Munich, Germany}
    
\author{Pascal~Weckesser}
    \affiliation{Max-Planck-Institut f\"{u}r Quantenoptik, 85748 Garching, Germany}
    \affiliation{Munich Center for Quantum Science and Technology (MCQST), 80799 Munich, Germany}

\author{Simon~Hollerith}
    \affiliation{Max-Planck-Institut f\"{u}r Quantenoptik, 85748 Garching, Germany}
    \affiliation{Munich Center for Quantum Science and Technology (MCQST), 80799 Munich, Germany}
    
\author{David~Wei}
    \affiliation{Max-Planck-Institut f\"{u}r Quantenoptik, 85748 Garching, Germany}
    \affiliation{Munich Center for Quantum Science and Technology (MCQST), 80799 Munich, Germany}

\author{Daniel~Adler}
    \affiliation{Max-Planck-Institut f\"{u}r Quantenoptik, 85748 Garching, Germany}
    \affiliation{Munich Center for Quantum Science and Technology (MCQST), 80799 Munich, Germany}
    \affiliation{Fakult\"{a}t f\"{u}r Physik, Ludwig-Maximilians-Universit\"{a}t, 80799 Munich, Germany}

\author{Immanuel~Bloch}
    \affiliation{Max-Planck-Institut f\"{u}r Quantenoptik, 85748 Garching, Germany}
    \affiliation{Munich Center for Quantum Science and Technology (MCQST), 80799 Munich, Germany}
    \affiliation{Fakult\"{a}t f\"{u}r Physik, Ludwig-Maximilians-Universit\"{a}t, 80799 Munich, Germany}

\author{Johannes~Zeiher}
    \affiliation{Max-Planck-Institut f\"{u}r Quantenoptik, 85748 Garching, Germany}
    \affiliation{Munich Center for Quantum Science and Technology (MCQST), 80799 Munich, Germany}

    
\date{\today}

%
\maketitle
\textbf{
Enhancing light-matter coupling at the level of single quanta is essential for numerous applications in quantum science.
The cooperative optical response of subwavelength atomic arrays has been found to open new pathways for such strong light-matter couplings, while simultaneously offering access to multiple spatial modes of the light field.
Efficient single-mode free-space coupling to such arrays has been reported, but the spatial control over the modes of outgoing light fields has remained elusive. 
Here, we demonstrate such spatial control over the optical response of an atomically thin mirror formed by a subwavelength array of atoms in free space using a single controlled ancilla atom excited to a Rydberg state.
The switching behavior is controlled by the admixture of a small Rydberg fraction to the atomic mirror, and consequently strong dipolar Rydberg interactions with the ancilla.
Driving Rabi oscillations on the ancilla atom, we demonstrate coherent control of the transmission and reflection of the array.
These results represent a step towards the realization of quantum coherent metasurfaces, the demonstration of controlled atom-photon entanglement and deterministic engineering of quantum states of light.
}

%
Realizing efficient light-matter interfaces and engineering states of light at the quantum level is challenging due to the small interaction cross section between atoms and photons~\cite{Chang2018}.
Overcoming this challenge requires enhanced coupling between light and matter, for example via optical cavities~\cite{Kimble1998, McKeever2004, Birnbaum2005, Mucke2010,Reiserer2015} or waveguides~\cite{Junge2013, Thompson2013, Lodahl2015, Lodahl2016} for single atoms, or by exploiting systems with high optical densities coupled to Rydberg states~\cite{Peyronel2012, Dudin2012, Firstenberg2016,Thompson2017}.
In optical cavities, for example, the presence or absence of a strongly coupled atom can be exploited to change the optical response of the cavity from transmitting to reflecting for impinging photons, the basis of photon-photon gates~\cite{Duan2004,Hacker2016,Stolz2022}. 
There, the enhanced interaction cross section, however, comes at the cost of a strong mode selection: optical cavities typically support only a single spatial mode for the photons, which limits their use for spatial light shaping.
Ordered subwavelength arrays of emitters have recently emerged as an alternative approach to realizing strong light-matter coupling~\cite{Porras2008,Jenkins2012,Jenkins2013,Facchinetti2016,Bettles2016,Shahmoon2017,Asenjo2017,Rui2020,Solntsev2021}, with distinct advantages over disordered ensembles in applications such as photon storage~\cite{Manzoni2018} or photonic gates~\cite{Bekenstein2020,Moreno2021}.
In these systems, emitters are periodically arranged at distances below the wavelength of light, resulting in highly cooperative optical properties as a result of dipolar interactions.
The free-space nature of cooperative arrays strongly relaxes the mode selection, which enables spatial control over the modes of single photons interacting with the array.
In particular, such control was recently proposed by using strong interactions between highly excited atomic Rydberg states~\cite{Bekenstein2020,Moreno2021,Zhang2022}.
The properties of a cooperative array can thereby be altered through the excitation of a single atom to a Rydberg state, realizing a ``quantum-controlled metamaterial" introduced by Bekenstein et al.~\cite{Bekenstein2020}, in which the optical response of the system can be changed in a spatially controlled way.
Such control provides a fundamentally new approach to photonic-state engineering in free space, with the perspective of creating large-scale photonic entangled states relevant for photonic quantum information applications~\cite{Bekenstein2020}.
Closely related proposals exploit a single atom to control the spatial photon mode via a dipole-dipole exchange interaction in an atomic ensemble~\cite{Petrosyan2018} or a bilayer atomic array~\cite{Grankin2018}.
In contrast to schemes based on dissipation~\cite{Peyronel2012, Dudin2012,Baur2014,Gorniaczyk2014,Firstenberg2016,Xu2021}, where decay occurs randomly from the input channel into a large number of modes by free-space scattering, the cooperative array allows for coherent switching between various spatial light modes, such as the transmission and reflection of an atomic array.
Furthermore, the cooperative optical response of ordered arrays dramatically reduces the required atom number and density to reach comparable optical depths as in disordered ensembles~\cite{Manzoni2018,Moreno2021}.
Similar to recent work performed in optical cavities~\cite{Stolz2022,Vaneecloo2022}, they can therefore help to mitigate known systematics that limit performance of disordered ensembles in free space at large atomic densities~\cite{Baur2014,Tiarks2019}.
\\
\begin{figure*}[t!]
    \centering
    \includegraphics{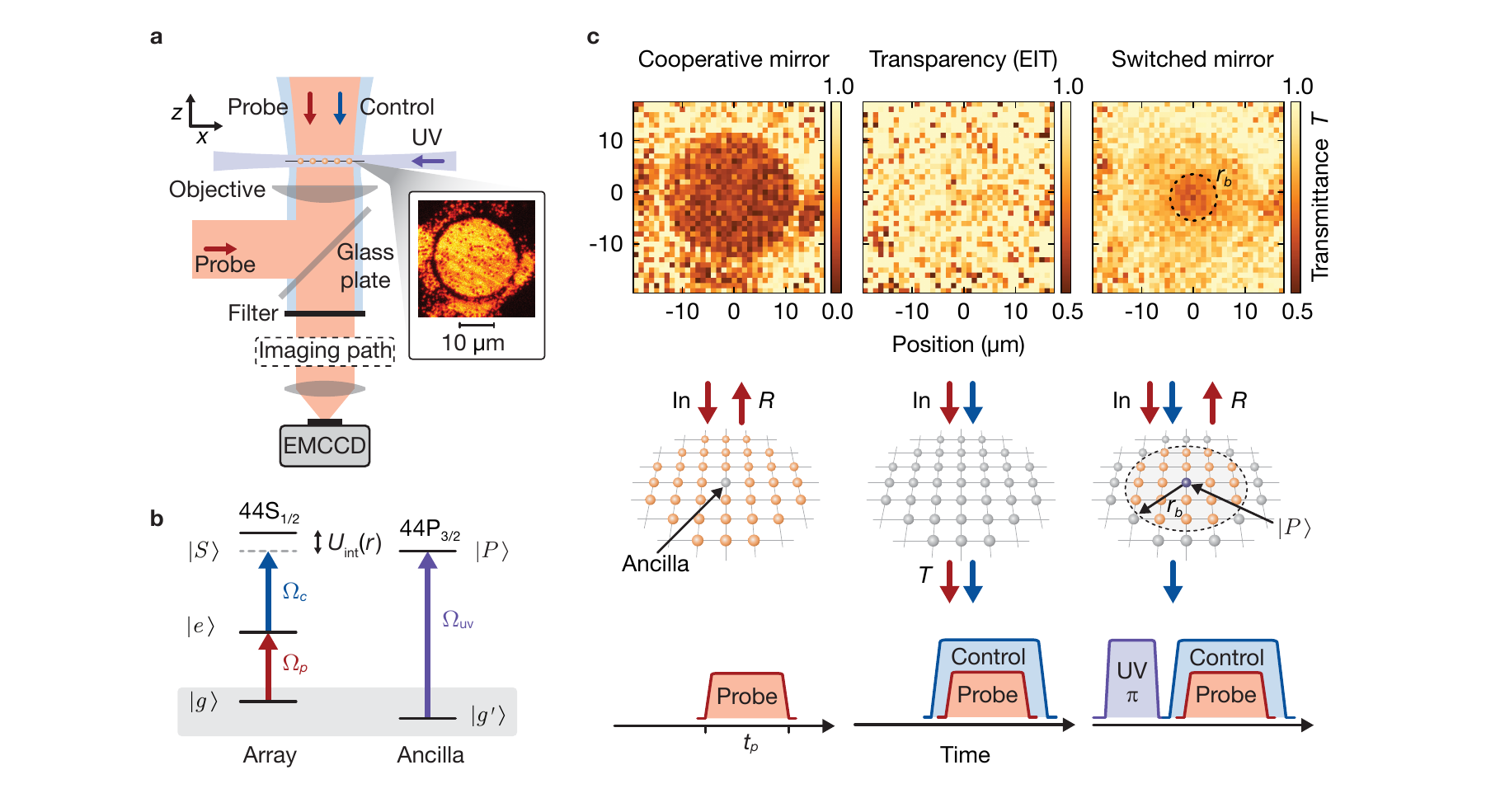}
    \caption{\textbf{Schematic of the experiment.}
        (\textbf{a}) Atomic array and laser beam orientations. The transmission (reflection) probe beam is overlapped and co- (counter-) propagating with the control beam along $-z$ ($+z$). We monitor the transmissive (reflective) response of the atomic array by imaging the probe beam onto an EMCCD, while filtering out the control beam.
        The atomic array is aligned in the $x-y$ plane, containing up to 1500 atoms in an atomic Mott insulator of a single atom per lattice site in state $\ket{g}$, while a single ancilla atom is prepared in a different hyperfine state $\ket{g'}$ at a target lattice site at the center of the array.
        We control the Rydberg excitation of the ancilla using an ultraviolet (UV) beam propagating in the atomic plane. The inset shows an exemplary site-resolved fluorescence image of a Mott insulator with 1500 atoms.
        (\textbf{b}) Electronic level scheme and relevant light fields. The control and probe fields with Rabi frequencies $\Omega_c$ and $\Omega_p$ respectively couple the ground state $\ket{g}$ to a Rydberg $S$-state $\ket{S}$ via an intermediate state $\ket{e}$. The UV field excites the ancilla $\ket{g'}$  with Rabi frequency $\Omega_{\mathrm{UV}}$ to a Rydberg $P$-state of the same principal quantum number $\ket{P}$. The $\ket{S}-\ket{P}$ Rydberg states experience a strong dipolar interaction, creating a distance-dependent shift in energy, $U_{\mathrm{int}}(r)$.
        (\textbf{c}) Spatially resolved optical response in transmission, the atomic array with the relevant light fields and the corresponding experimental pulse sequences.
        (Left) With the probe field alone, the atomic array acts as a cooperative mirror. (Middle) Applying an additional resonant control field, we render the atomic array transparent exploiting the EIT condition. (Right) Preparing the ancilla in the $\ket{P}$ state, the dipolar Rydberg interaction shifts the control field out of resonance, restoring the reflectivity within a finite radius around the ancilla. The dashed line indicates the estimated blockade radius of $r_b = \SI{4.6}{\micro\meter}$~\cite{SI}.
    }
    \label{fig:1}
\end{figure*}
%
Here, we exploit the strong cooperative response of an array of ordered emitters separated by subwavelength distances to realize a switch for photons.
This setup allows for recreating the prototypical situation encountered in strongly coupled cavity-QED, where single-atom control can be exploited to reroute single photons~\cite{Reiserer2015}.
We utilize the strong interactions between Rydberg states of opposite parity to switch the optical properties of the array from transmitting to reflecting.
We achieve spatial control by using an ancilla atom prepared with single-site precision at a specific target position within the array.
We demonstrate that the optical properties of the array can be altered coherently by driving Rabi oscillations on the ancilla into the Rydberg state.
Finally, we directly measure the spatial switching area of the ancilla in our system and present evidence that the residual imperfections in switching are dominated by the finite Rydberg lifetime and preparation fidelity of the ancilla, both straightforward to overcome with future upgrades to the experimental setup.

Analogous to recent experiments focused on quantum optics with Rydberg atoms~\cite{Peyronel2012,Baur2014,Firstenberg2016}, the key idea for controlling our subwavelength array is to transfer the strong interactions between Rydberg states to the optical response of the cooperative array through electromagnetically induced transparency (EIT)~\cite{Fleischhauer2005}.
We start with a cooperative atomic array with emitters approximately described as two-level systems with ground state $\ket{g}$ and excited state $\ket{e}$.
To induce EIT, the excited state $\ket{e}$ is coupled with a control field $\Omega_c$ to a highly excited Rydberg $S$-state $\ket{S}$ (see Fig.~\ref{fig:1}b).
As a result, the cooperative optical two-level response for a weak probe field of Rabi frequency $\Omega_p$, impinging normal on the array is, altered and the system becomes transparent on the $\ket{g}\leftrightarrow \ket{e}$ resonance in presence of the control beam (middle column in Fig.~\ref{fig:1}c).
In establishing transparency, the excited state $\ket{S}$ is admixed to the state $\ket{e}$ through the control field $\Omega_c$.
Consequently, $\ket{e}$ inherits some of the long-range interacting character of $\ket{S}$.
The parameters $\Omega_c$ and $\Omega_p$ are chosen to keep the Rydberg state population sufficiently small to avoid optical nonlinearities due to self-blockade, which is expected when the probability to find any array atom in $\ket{S}$ approaches unity~\cite{Pritchard2010, Peyronel2012}.
To control the properties of the cooperative mirror, an additional ``ancilla" atom in the ground state $\ket{g'}$ is excited to a neighboring Rydberg $P$-state $\ket{P}$.
\begin{figure*}[t!]
    \centering
    \includegraphics{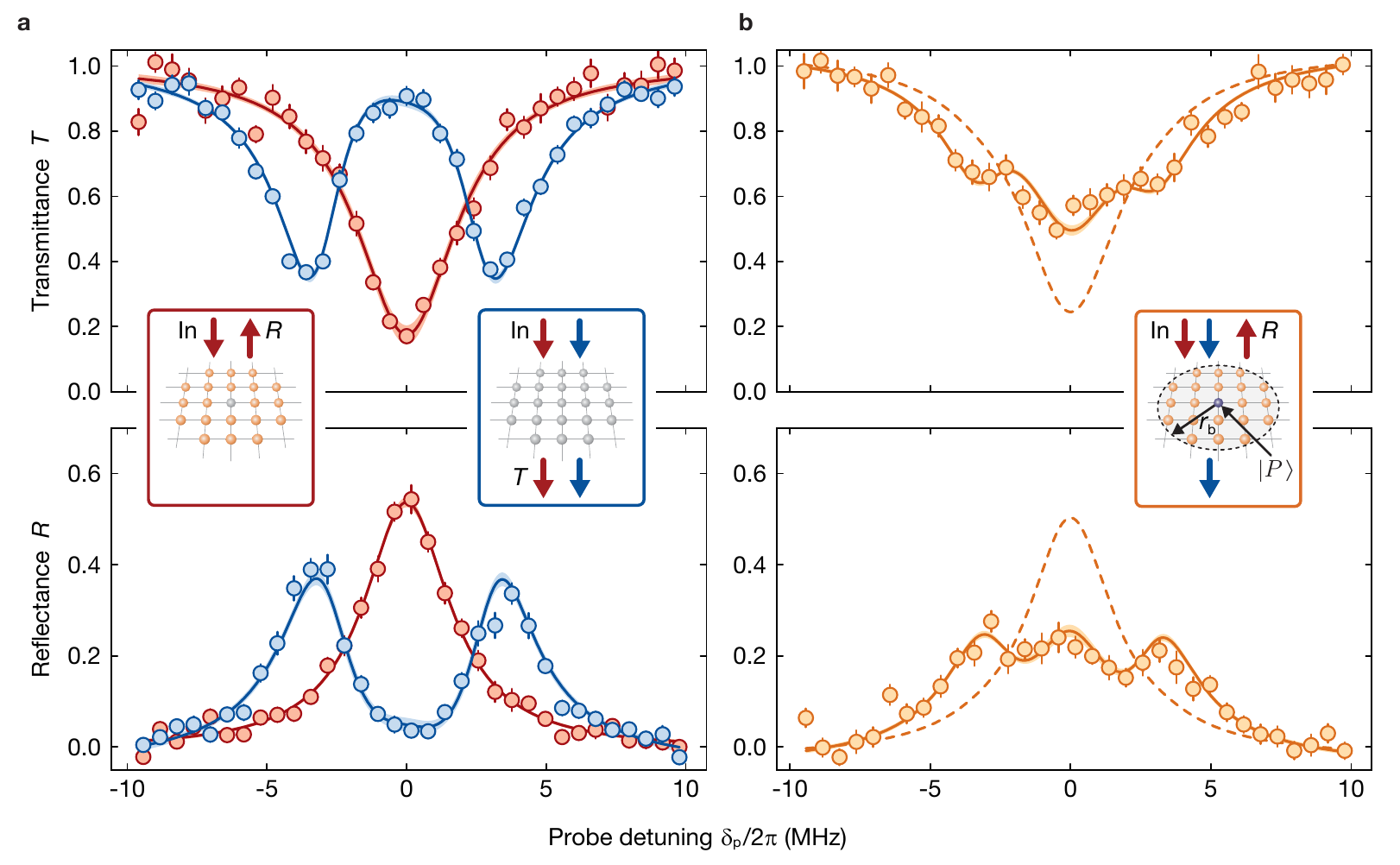}
    \caption{\textbf{Cooperative response in absence and presence of the Rydberg ancilla.}
    (\textbf{a}) Cooperative response of the atomic array with (blue) and without (red) control beam for a probe duration of $t_p = \SI{20}{\micro\second}$ and the ancilla prepared in $\ket{g'}$.
    Without the control laser, we reproduce the cooperative subradiant response of the mirror with a transmission\,(reflection) linewidth of $\Gamma_{\mathrm{M}}/2\pi = 4.40(32) \,(3.75(14))\,\si{\mega\hertz}$.
    With the control laser present, we observe a splitting of the single peak into an EIT doublet, where the width of each peak amounts to $\Gamma_{\mathrm{EIT}}/2\pi = 2.95(17)\,(3.01(28))\, \si{\mega\hertz}$ in transmission\,(reflection) and a minimal (maximal) transmittance\,(reflectance) of $0.35(2)\,(0.37(2))$ is observed.
    (\textbf{b}) Preparing the ancilla atom in the Rydberg state $\ket{P}$, the spectra (orange) change dramatically, and reveal a triple-peak structure, featuring contributions of both the cooperative mirror and EIT spectrum.
    Superimposing both spectra while having the ancilla Rydberg fraction $P_{\ket{P}}$ and a global offset as free fit parameters, we find excellent agreement with our data set with $P_{\ket{P}} = 0.61(2)\,(0.45(2))$ in transmittance\,(reflectance), in good agreement with an independent reference measurement of $P_{\ket{P}} = 0.52(8)$~\cite{SI}.
    The dashed lines illustrate the expected spectra assuming ideal ancilla preparation and substantially shorter probe duration than the Rydberg lifetime ($t_p = \SI{2}{\micro\second}$), improving the ancilla Rydberg fraction to $P_{\ket{P}} = 0.96$.
    The insets in each figure illustrate the atomic array and beam directions for each experimental configuration.
    The red (blue) arrows indicate the incident and scattered probe (control) beam directions.
    The measurements are averages over $70-125$ independent repetitions.
    Error bars denote the standard error of the mean (s.e.m.).
    }
    \label{fig:2}
\end{figure*}
Due to strong Rydberg interactions between $\ket{S}$ and $\ket{P}$, the state $\ket{S}$ is shifted in energy by $U_{\mathrm{int}}(r)$.
This interaction shift exceeds half of the EIT spectral width within a ``blockade disc" of radius $r_b$~\cite{Gunter2012,Gunter2013} centered around the location of the ancilla.
Consequently, the EIT condition breaks down and the optical properties return to those of the cooperative mirror.
%
Due to its coherent nature, the single ancilla atom can entangle the mirror response with the ancilla state, which can subsequently be exploited for photonic-state engineering~\cite{Bekenstein2020}.
Furthermore, controlling the position of the ancilla atom within the array enables full spatial control over the optical properties of the array (see Fig.~\ref{fig:1}c).
In particular, using this scheme, optical modes with diameters of few lattice sites can be controlled without compromising the cooperativity of the response~\cite{Manzoni2018}.

We began our experiments by preparing a nearly unity filled two-dimensional atomic array of $^{87}\mathrm{Rb}$ atoms spin-polarized in the state ${\ket{g} = \ket{5\mathrm{S}_{1/2},F=2,m_F=-2}}$ in a single vertical antinode of a three-dimensional optical lattice with lattice constant $a_{\mathrm{lat}}=\SI{532}{\nano\meter}$.
The lattice spacing was below the transition wavelength ${\lambda_p = \SI{780}{\nano\meter}}$ from the ground state $\ket{g}$ to the excited state ${\ket{e} =  \ket{5\mathrm{P}_{3/2},F=3,m_F=-3}}$, leading to a cooperative response of the array at a ratio of ${a_{\mathrm{lat}}/\lambda_{p} = 0.68}$~\cite{Rui2020}.
To enable control of the optical response of the mirror via Rydberg interactions, we coupled the excited state $\ket{e}$ to the $\ket{S} = \ket{44\mathrm{S}_{1/2},m_J = -1/2}$ Rydberg state.
The optical properties of the array were probed with a weak probe beam with Rabi coupling $\Omega_p/2\pi = \SI{168(5)}{\kilo\hertz}\ll\Omega_c/2\pi = \SI{6.7(6)}{\mega\hertz}$ on the $\ket{g}\leftrightarrow \ket{e}$ transition, which is sufficient to create the EIT window and admix a small Rydberg amplitude to the excited state $\ket{e}$~\cite{Fleischhauer2005}. 
We deterministically created a single ancilla atom in the ${\ket{g'} = \ket{5\mathrm{S}_{1/2},F=1,m_F=-1}}$ state with a fidelity of $0.83(4)$ at the center of the array using single-site addressing~\cite{Weitenberg2011,Fukuhara2013}.
The ancilla was then controllably excited to the Rydberg state $\ket{P} = \ket{44\mathrm{P}_{3/2},m_J = 3/2}$ on an ultraviolet (UV) transition at a wavelength of $\SI{297}{\nano\meter}$. 
The interaction with the admixed $\ket{S}$-state Rydberg fraction of the array atoms led to a Förster-enhanced energy shift $U_{\mathrm{int}}(r)$, featuring a characteristic van der Waals distance dependence proportional to ${C_6/r^6}$ due to interactions with nearby Zeeman sublevels~\cite{SI}.
For our parameters, this resulted in a blockade radius ${r_b = (2C_6\Gamma_e/\Omega_c^2)^{1/6}=\SI{4.6}{\micro \meter}}$~\cite{SI} which defined the range over which the mirror properties are altered.
Preparing an atomic array with a radius $r_a\gg r_b$ and detecting the probe light with a low-noise electron-multiplying change-coupled device (EMCCD) camera, we directly reveal the spatially switched area and demonstrate the spatially selective response of our array, see Fig.~\ref{fig:1}c.
To suppress the effect of long-range dipolar exchange of the ancilla~\cite{Gunter2013,Schempp2015}, we work in a regime $r_a\approx r_b$ for the following characterization of the switching response of the array~\cite{SI}.

In a first set of experiments, we aim at demonstrating the basic mechanism of switching the cooperative mirror by the ancilla atom.
To this end, we first confirm the cooperative nature of our atomic array by measuring the reflection and transmission response of a laser beam tuned near the resonance of the $\ket{g} \leftrightarrow \ket{e}$ transition. 
We find a strong directional signal with a subradiant Lorentzian lineshape and an extracted width of down to $\Gamma_{\mathrm{M}}/2\pi = \SI{3.75(14)}{\mega\hertz}$, narrower than the natural linewidth of $\Gamma_e/2\pi = \SI{6.06}{\mega\hertz}$, see Fig.~\ref{fig:2}a.
This confirms that our array is in the cooperative regime explored previously~\cite{Rui2020}.
Illuminating the array with both the probe and control fields on resonance, we observe EIT, resulting in a switching from a reflecting to transmitting atom array.
Scanning the probe detuning we observe that the dip in transmission (peak in reflection) splits into a doublet with splitting $\Omega_c$, demonstrating the effects of the control field.
Interestingly, this doublet again shows signatures of a cooperative response, with a high level of reflectance of $0.37(2)$, exceeding the reflectance signal for isotropic scattering 0.16(3)~\cite{SI}.
The width of each peak amounts to $\Gamma_{\mathrm{EIT}}/2\pi = \SI{2.95(17)}{\mega\hertz}$, consistent with the width in the single-particle limit of $\Gamma_e/2$.
Our parameters were chosen to maximize the on-resonance contrast between the cooperative mirror and the EIT response using a probe duration $t_p = \SI{20}{\micro\second}$, only slightly lower than the measured lifetime $\tau = \SI{27(5)}{\micro\second}$ of the Rydberg ancilla. 
The probe power and pulse duration were chosen to keep the Rydberg admixture and the effects of self-blockade small, while providing sufficient signal-to-noise ratio of the probe light on the EMCCD camera~\cite{SI}.\\
To investigate the effect of the Rydberg ancilla on the array, we apply a $\pi$-pulse on the $\ket{g'} \leftrightarrow \ket{P}$ transition with duration $t$, such that $\Omega_{\mathrm{UV}}t = \pi$. 
The resulting spectra exhibit a broad resonance featuring a substructure of three distinct peaks, with the reflectance on resonance amounting to $0.25(2)$, see Fig.~\ref{fig:2}b.
Our observation of a triple peak structure can be understood to arise from a combination of the configurations with and without the ancilla Rydberg atom present.
A simplified model assuming a statistical mixture of the mirror in the switched and unswitched state, weighted with the probability of finding the ancilla in the Rydberg or ground state respectively, quantitatively reproduces the observed features in ~Fig.~\ref{fig:2}b.
The ansatz of a statistical mixture of the two mirror states is motivated by imperfect initial state preparation of the ancilla in $\ket{g'}$ and decay of the ancilla Rydberg state during probing~\cite{SI}.
To illustrate potential improvements in an upgraded experimental setup, the dashed lines in Fig.~\ref{fig:2}b also show the expected spectra for perfect ancilla preparation in $\ket{g'}$ and substantially shorter probe duration of $t_p = \SI{2}{\micro\second}$, for which on the order of one photon is scattered, and the decay of the Rydberg ancilla becomes negligible.

\begin{figure}[t!]
    \centering
    \includegraphics{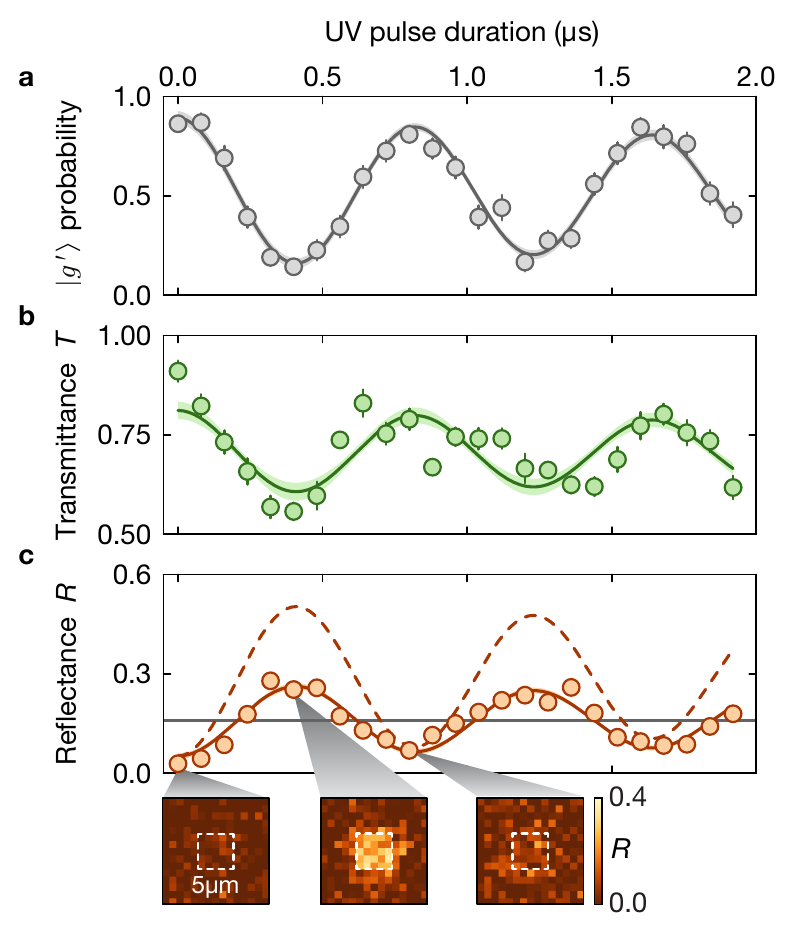}
    \caption{\textbf{Cooperative response after a coherent drive of the ancilla.}
    (\textbf{a}) Ancilla $\ket{g'}\leftrightarrow \ket{P}$ Rabi oscillations obtained from ground state $\ket{g'}$ fluorescence detection by varying the length of the UV pulse before probing, see Fig.~\ref{fig:1}c for the protocol.
    Applying a damped sinusoidal fit, we find a Rabi frequency of $\Omega_{\mathrm{UV}}/2\pi = \SI{1.22(2)}{\mega\hertz}$ and a decay constant of $\tau_{\mathrm{decay}} = \SI{6(3)}{\micro\second}$. 
    We observe the transmittance (\textbf{b}) and reflectance (\textbf{c}) data to follow the Rabi oscillation of the ancilla.
    The solid lines in \textbf{b} and \textbf{c} represent the best fit results, with the amplitude of the oscillation and overall offset as the only fit parameters, while the oscillation frequency and decay time are fixed and taken from \textbf{a}.
    The three insets in \textbf{c} indicate spatially-averaged reflection images for $\Omega_{\mathrm{UV}}t = 0$, $\pi$ and $2\pi$, respectively, with an indicated region-of-interest (ROI) of $5\times5\,\si{\micro\meter}$.
    The dashed line in \textbf{c} illustrates the expected transmission signal for an ancilla Rydberg fraction of $P_{\ket{P}} = 0.96$.
    The gray solid line represents the resonance reflection signal ($0.16(3)$) from isotropic scattering.
    The latter was measured experimentally by introducing vertical disorder by means of Bloch oscillations~\cite{SI}.
    The measurements are averages over $120-170$ independent repetitions.
    Error bars denote the s.e.m..
    }
    \label{fig:3}
\end{figure}

\begin{figure}[t!]
    \centering
    \includegraphics{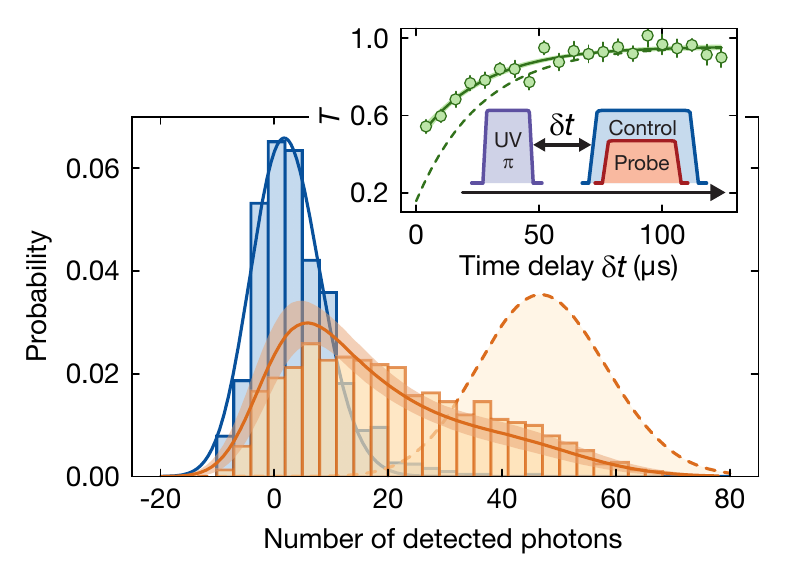}
    \caption{\textbf{Distribution of detected photon number and lifetime.}
        Detected photon number distribution, relative to the mean background photon number, in reflection within a ROI by either preparing the ancilla in the $\ket{g'}$ (blue) or $\ket{P}$ (orange) state.
        For the former, we obtain a Poissonian distribution ($N=100$ repetitions) corresponding to the photon counts in the EIT configuration (blue solid line).
        Preparing the ancilla in $\ket{P}$, the histogram acquires a tail towards high reflected photon numbers ($N = 130$ repetitions).
        This histogram is a combination of counts due to Rydberg-induced reflection (orange dashed line), and counts at low photon numbers due to imperfect Rydberg preparation and Rydberg decay.
        A Monte-Carlo simulation including our experimental uncertainties reproduces the essential features of the observed histogram (orange solid line and shaded region; see~\cite{SI}).
        The inset displays the transmission signal for variable delay time $\delta t$ between Rydberg excitation and probe pulse ($70$ repetitions).
        We extract the Rydberg lifetime $\tau = \SI{27(5)}{\micro\second}$ from an exponential fit (solid line).
        The green dashed line is the expected transmission signal for an ancilla Rydberg fraction of $P_{\ket{P}} = 0.96$.
        Error bars denote the s.e.m..
    }
    \label{fig:4}
\end{figure}
To highlight the capability of coherent manipulation in our system, we next aim to dynamically change the optical properties of the atomic array.
To this end, we drive the ancilla from the ground state $\ket{g'}$ to the Rydberg state $\ket{P}$ with variable UV pulse durations, resulting in coherent Rabi oscillations of the ancilla with a Rabi frequency $\Omega_{\mathrm{UV}}/2\pi = \SI{1.22(2)}{\mega\hertz}$, see Fig.~\ref{fig:3}.
Measuring the transmittance or reflectance of the array in the same sequence, we find a strong correlation between ancilla Rabi oscillations and the optical properties of the array, where the mirror switches from transmitting to reflecting during the course of the oscillations.
Fitting the dynamics of transmittance and reflectance of the array with a damped sinusoidal function derived from the Rabi oscillations with the amplitude and offset of the oscillation as free parameters, we find excellent agreement between this model and the data.
This agreement indicates that, indeed, the switching behavior is determined by the quantum state of the ancilla before probing.
The small distortions in the transmittance can be attributed to a non-vanishing probability to initially have two ancilla atoms before excitation to the Rydberg state~\cite{SI}. 
Notably, the maxima of the oscillating reflectance are clearly above the single-particle limit of a vertically disordered array~\cite{SI,Rui2020}, demonstrating that the cooperative response of the mirror is preserved during the oscillation, see Fig.~\ref{fig:3}c.

The strong correlation between the state of the ancilla and the state of the mirror can be further studied through photon number statistics.
In the ideal case, we expect all photons within a detection window to be reflected (transmitted) when the ancilla is excited to the Rydberg state $\ket{P}$ (in its ground state $\ket{g'}$).
We study this correlation by monitoring the number of reflected photons for a longer integration time of $t_p = \SI{60}{\micro\second}$ after controllably exciting the ancilla with a $\pi$-pulse, see Fig.~\ref{fig:4}.
The distribution with the ancilla in $\ket{P}$ exhibits a long tail at high numbers of reflected photons in addition to a peak at low photon numbers.
We find good agreement of our observed histogram with a model taking into account our estimated preparation fidelity as well as the independently measured lifetime of the Rydberg-excited ancilla via Monte-Carlo sampling~\cite{SI}, see Fig.~\ref{fig:4}.

\begin{figure}[t!]
    \centering
    \includegraphics{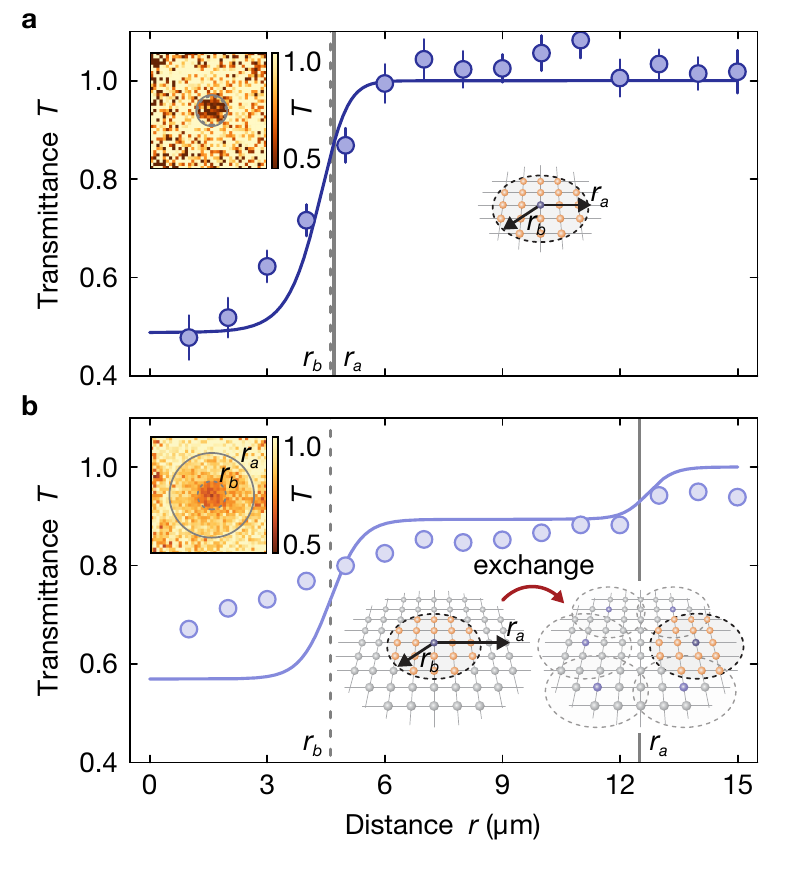}
    \caption{\textbf{Spatially resolved switching area.}
     Radially averaged transmittance over the size of the switched mirror, centered around the ancilla atom.
     The solid lines show the estimated radial profile, including $\ket{S}-\ket{P}$ blockade, array size and Rydberg fraction of the ancilla $P_{\ket{P}}$.
     (\textbf{a}) The transmittance of the array containing $250$ atoms, comparable in size ($r_a\approx\SI{4.7}{\micro\meter}$) with the blockade radius, shows good agreement with the estimated radial profile ($120$ repetitions).
     (\textbf{b}) The array of $1500$ atoms with the radius $r_a\approx\SI{12.5}{\micro\meter}$, large compared to blockade radius, shows a washing out of the transmittance versus distance, deviating from the solid line ($3500$ repetitions).
     This can be explained by long-range exchange processes, where the $\ket{P}$ excitation undergoes $\ket{S}-\ket{P}$ exchange resulting in the transport of the excitation and therefore a shift of the switching area as illustrated by the sketch.
     The insets on the upper left illustrate the spatially-averaged transmission images, respectively.
     The vertical solid (dashed) lines mark the array radius $r_a$ (estimated blockade radius $r_b$).
     Error bars denote the s.e.m..
    }
    \label{fig:5}
\end{figure}
The spatial control over the position of the ancilla allows for a fundamentally new approach to controlling the optical response of the subwavelength array in a spatially resolved way.
To demonstrate such control, we prepared the ancilla at a target site in the center of the array and compared the optical response of a small array of radius $r_a=\SI{4.7(7)}{\micro\meter}$ with a larger array with a radius $r_a=\SI{12.5(5)}{\micro\meter}$, which exceeds the expected blockade radius, see Fig.~\ref{fig:5}.
In the small array, we observe a relatively sharp edge where the transmission jumps from its central value of $0.48(2)$ to near unity, due to the combination of the finite size of the array $r_a$ and the blockade radius $r_b$.
In contrast, the large array has an increased transmittance at the center as well as a more gradual increase of the transmittance beyond the blockade radius.
These observations indicate the presence of previously studied long-range exchange processes~\cite{Gunter2013,Schempp2015}, which cause the ancilla to delocalize over the entire system and lead to a smoothened transmission signal.
Importantly, these exchange processes can be suppressed by either operating on shorter probe timescales or by reducing the probe power, as the relevant exchange process scales with $\propto \Omega_p^2$~\cite{SI,Schempp2015}.
This regime is experimentally accessible with optimized detectors matched to the manipulated spatial modes of the light field, which, however, would not have allowed for the spatially resolved proof-of-principle characterization of the array response performed in this work.

In conclusion, we have demonstrated the ability to switch and coherently control the optical properties of a cooperative subwavelength array of atoms using a single ancilla atom.
Our system is presently limited by finite preparation efficiencies as well as the finite Rydberg lifetime of the ancilla.
The former can be improved by better addressing techniques, e.g., by placing the ancilla in a single microtrap overlapped with the cooperative array, the latter with optimized single-photon detectors.
Alternatively, we foresee the use of Rydberg dressing~\cite{Jau2016} for the ancilla, improving its lifetime while minimizing motional decoherence effects due to reduced repulsion in the lattice and the suppression of dipolar exchange processes.
Our measurements already demonstrate all experimental building blocks to control single photons by manipulating single atoms in subwavelength arrays, and open the path towards the detection of atom-photon entanglement~\cite{Li2013}, the realization of photon-photon gates~\cite{Bekenstein2020, Gorshkov2011} or multimode quantum optics in cooperative arrays~\cite{Bekenstein2020,Moreno2021,Zhang2022}.

\begin{acknowledgments}
We gratefully acknowledge discussions with Ignacio Cirac, Rivka Bekenstein, Jun Rui, Darrick Chang, Ephraim Shahmoon and Sebastian Weber.
We acknowledge funding by the Max Planck Society (MPG) and from Deutsche Forschungsgemeinschaft (DFG, German Research Foundation) under Germany’s Excellence Strategy – EXC-2111 – 390814868 and Project No. BL 574/15-1 within SPP 1929 (GiRyd).
This project has received funding from the European Union’s Horizon 2020 research and innovation programme under grant agreement No. 817482 (PASQuanS).
J.Z. acknowledges support from the BMBF through the program ``Quantum technologies - from basic research to market" (Grant No. 13N16265).
K.S. acknowledges funding through a stipend from the International Max Planck Research School (IMPRS) for Quantum Science and Technology.
\end{acknowledgments}

\textbf{Author Contribution Statement:}
K.S. acquired the data and, together with P.W., D.W. and D.A., maintained and improved the experimental setup.
P.W. and S.H. contributed the theoretical simulations.
I.B. and J.Z. supervised the study.
All authors worked on the interpretation of the data and contributed to the final manuscript.

\textbf{Competing interests:} The authors declare no competing interests.

\textbf{Accepted Manuscript:} This version of the article has been accepted for publication, after peer review (when applicable)
but is not the Version of Record and does not reflect post-acceptance improvements, or any
corrections. The Version of Record is available online at: \url{https://doi.org/10.1038/s41567-023-01959-y}.


\bibliography{rydberg_mirror}
\clearpage


\setcounter{equation}{0}
\setcounter{figure}{0}
\setcounter{table}{0}
\renewcommand{\theequation}{S\arabic{equation}}
\renewcommand{\theHequation}{S\arabic{equation}}
\renewcommand{\thefigure}{S\arabic{figure}}
\renewcommand{\theHfigure}{S\arabic{figure}}
\renewcommand{\thetable}{S\arabic{table}}
\renewcommand{\theHtable}{S\arabic{table}}

\section*{Supplementary Information}

\section{Experimental details}
In this section, we give a detailed description of the experimental methods including initial state preparation, experimental sequence, detection of the probe field, and Rydberg excitation scheme.
\subsection{Initial state preparation and experimental sequence}
\begin{figure}[t!]
    \centering
    \includegraphics{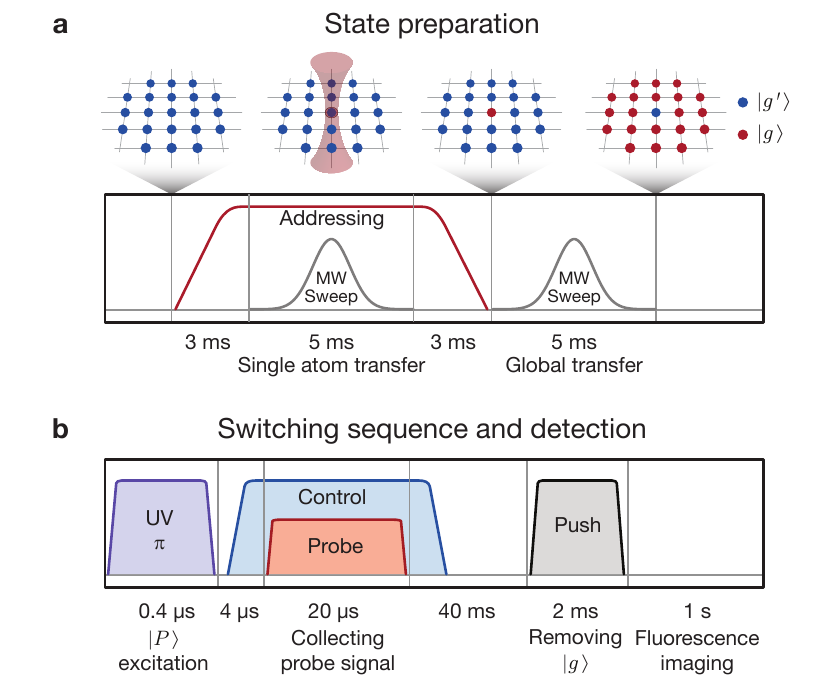}
    \caption{\textbf{State preparation and experimental sequence.}
        (\textbf{a}) Starting with a near unity-filled Mott insulator in $\ket{g'}$, we use a microwave (MW) transfer combined with a tightly focused addressing beam at a wavelength of $\SI{787.55}{\nano\meter}$ to transfer a single ancilla atom to $\ket{g}$.
        Then, a subsequent global MW sweep prepares the ancilla atom in $\ket{g'}$ and the array atoms in $\ket{g}$.
        (\textbf{b}) Switching of the cooperative mirror is then performed by exciting of the ancilla to $\ket{P}$ with an ultraviolet (UV) $\pi$-pulse, $\Omega_{\mathrm{UV}}t=\pi$.
        After switching on the control beam and an additional waiting time of $\SI{4}{\micro\second}$, the probe beam is switched afterwards to monitor the probe signal on the EMCCD camera.
        Depending on the state to be detected, a resonant push-out can be added to remove $\ket{g}$ before recording the fluorescence image.
        }
    \label{fig:s1}
\end{figure}

Our experiment started with a 2D Bose-Einstein condensate of $^{87}\mathrm{Rb}$ atoms, confined in a single antinode of a vertical ($z$-axis) optical lattice.
By adiabatically ramping up two lattices along the $x$ and $y$ direction with lattice constant $a_{\mathrm{lat}}=\SI{532}{\nano\metre}$, we created a near unity-filled Mott insulator, with $\approx250$\,(1500) atoms and a filling of $\eta \approx 0.96\,(0.92)$ ~\cite{Wei2022}.
Exploiting the single-site resolution of our quantum gas microscope, we deterministically prepared a single atom in the ground state Zeeman sublevel $\ket{g'} = \ket{5P_{1/2},F=1,m_F=-1}$ at the center of the cloud~\cite{Fukuhara2013,Weitenberg2011}, whereas the atoms forming the subwavelength array remained in $\ket{g} = \ket{5P_{1/2},F=2,m_F=-2}$, see Fig.~\ref{fig:s1}a.\\
To reduce the spatial wavefunction spread of the atoms within the array, we ramped up all three lattices to a depth of $100\,E_r$, where $E_r = h^2/8ma_{\mathrm{lat}}^2$ is the recoil energy of the lattice.
Afterwards, we applied an excitation pulse in the ultraviolet spectral range (UV) with an area of $\Omega_{\mathrm{UV}} t = \pi$ to controllably excite the ancilla atom to the Rydberg state.
With the ancilla prepared, we turned on the control beam $\SI{4}{\micro\second}$ before the probe beam, which was kept on for a duration of $t_p = \SI{20}{\micro\second}$, and collected the transmission (reflection) signal on the EMCCD camera.
After that, we performed site-resolved fluorescence imaging with our quantum gas microscope~\cite{Sherson2010} to obtain the lattice occupation in addition to the reflectance/transmittance of the array.
For the Rabi oscillations shown in Fig.~3, the state of the ancilla is detected via loss of the Rydberg state~\ket{P} from the optical lattice, such that the absence of the ancilla in the final fluorescence image signals the excitation to the Rydberg state $\ket{P}$.
To distinguish the ancilla in state $\ket{g'}$ from the surrounding ground state atoms $\ket{g}$, we applied a resonant push-out pulse on the latter before taking the fluorescence image, see Fig.~\ref{fig:s1}b.

\subsection{Detection of the probe field}
The probe beam and the control beam propagated perpendicular to the atomic plane, see Fig.~1a.
The optical field at the position of the array was imaged by a high-resolution objective with a numerical aperture (NA) of $0.68$ onto an electron multiplying charge-coupled device (EMCCD) camera (iXon Ultra 897, Andor Technology), which has a quantum efficiency of $\SI{80}{\percent}$.
The control beam was filtered out before the EMCCD camera to suppress background for the detection of the $\SI{780}{\nano\meter}$ probe light.\\
In the experiment, the probe beam parameters are constrained by three important factors.
First, the finite lifetime of the ancilla Rydberg state limits the maximum probe duration of the switched array.
Second, high probe powers lead to a Rydberg self-blockade for the array during EIT probing when the admixed Rydberg fraction per blockade volume exceeds unity, see Sec.~\ref{Nonlinearity}.
This bounds the incident probe photon Rabi frequency $\Omega_p$ for a given control Rabi frequency $\Omega_c$.
Third, the number of photons incident on the EMCCD camera has to be sufficient to overcome detection noise.
To strike a compromise between all criteria, we first set the probe duration to $t_p = \SI{20}{\micro\second}$.
Then, the probe Rabi coupling $\Omega_p$ was reduced to minimize self-blockade, while still obtaining enough signal-to-noise on the EMCCD camera.
For the transmission beam, a direct measurement with the EMCCD yielded about $0.36(2)$ photons incident on one lattice site during a duration of $t_p = \SI{20}{\micro\second}$, which corresponds to $\Omega_p/2\pi = \SI{168(5)}{\kilo\hertz}$.\\
For the reflection measurement, the probe beam was combined with the imaging path using a glass plate, reflected through the objective and focused on the atoms.
We calibrated the incident intensity of the reflection probe beam by comparing the resonant atomic heating due to the reflection probe with that of the transmission probe in shallow lattices~\cite{Rui2020}.
The estimated incident photon flux for all reflectance measurements is $0.44(8)$ photons per lattice site within a duration of $t_p = \SI{20}{\micro\second}$, which corresponds to $\Omega_p/2\pi = \SI{189(16)}{\kilo\hertz}$.

\begin{figure}[t!]
    \centering
    \includegraphics{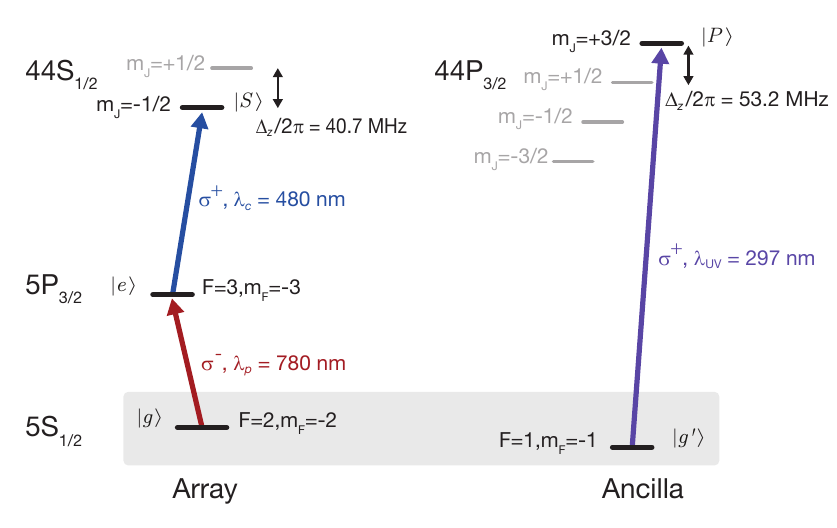}
    \caption{\textbf{Electronic level structure and excitation scheme.}
        The array atoms are resonantly coupled to Rydberg states through a two-photon transition from ${\ket{5S_{1/2},F=2,m_F=-2}}$ to ${\ket{44S_{1/2},m_j=-1/2}}$ via the intermediate state $\ket{5P_{3/2},F=3,m_F=-3}$ with $\sigma^-$\,($\sigma^+$) polarized probe\,(control) beam.
        The ancilla atom, initially prepared in ${\ket{5S_{1/2},F=1,m_F=-1}}$, is coupled to ${\ket{44P_{3/2}, m_j=+3/2}}$ using the $\sigma^+$ polarized component of the UV laser.
        }
    \label{fig:s2}
\end{figure}

%
\subsection{Excitation scheme}
Fig.~\ref{fig:s2} shows the relevant electronic states of the array atoms and the ancilla atom including the respective transitions.
The array atoms were coupled from ${\ket{g} = \ket{5S_{1/2},F=2,m_F=-2}}$ to ${\ket{e} = \ket{5P_{3/2},F=3,m_F=-3}}$ with the probe beam at a wavelength of $\lambda_p = \SI{780}{\nano\meter}$ and from $\ket{e}$ to ${\ket{S} = \ket{44S_{1/2},m_J=-1/2}}$ with the control beam at a wavelength of $\lambda_c = \SI{480}{\nano\meter}$.
We applied a magnetic field of $B_z = 28.5\,G$ perpendicular to the atomic plane to isolate different Zeeman sublevels.
Both probe and control beams propagated parallel to the bias field, setting a quantization axis for the almost pure $\sigma^-$ or $\sigma^+$ polarization in both transitions.
The ultraviolet (UV) excitation pulse at a wavelength of ${\lambda_{\mathrm{UV}} = \SI{297}{\nano\meter}}$ coupled the ancilla atom, which was prepared in the ground state ${\ket{g'} = \ket{5S_{1/2},F=1,m_F=-1}}$, to the state ${\ket{P} = \ket{44P_{3/2},m_J=+3/2}}$ without affecting the array atoms in $\ket{g}$.
A large Zeeman splitting of ${\Delta_z/2\pi = \SI{53.2}{\mega\hertz}}$ within the $44P_{3/2}$ manifold isolates $\ket{P}$ state of the ancilla and relaxes the requirements on polarization purity of the UV transition.


\section{Rydberg interactions}
In this section, we discuss the influence of self-blockade, the details of the $\ket{S}-\ket{P}$ Rydberg interaction, the EIT-blockade radius, and provide a quantitative derivation of the $\ket{S}-\ket{P}$ exchange process.
%
%
\subsection{Optical nonlinearity due to self-blockade}
\label{Nonlinearity}
\begin{figure*}[t!]
    \centering
    \includegraphics{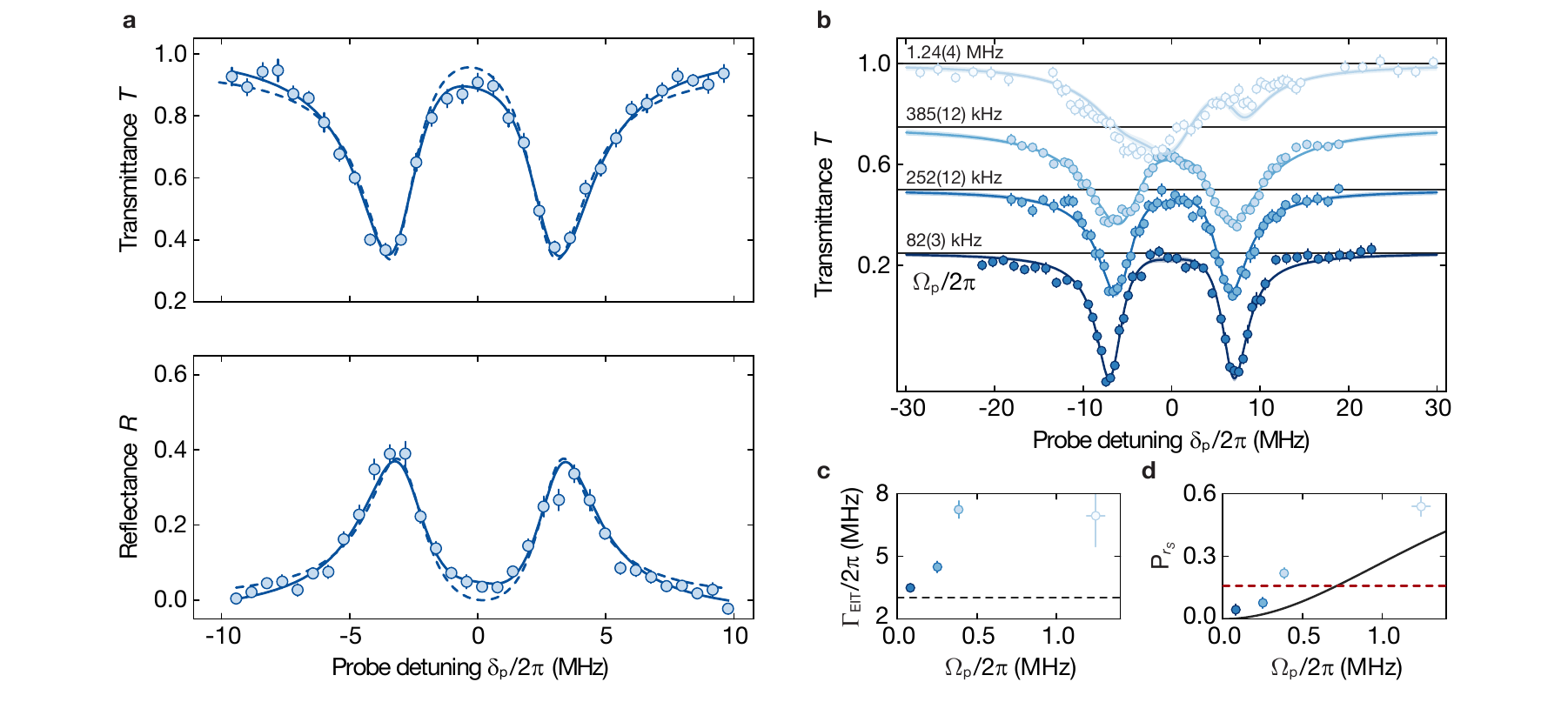}
    \caption{\textbf{EIT nonlinearity due to Rydberg self-blockade.}
        (\textbf{a}) Standard single-atom EIT model in Eq.~\ref{eq:s0} (dashed line) and Rydberg EIT model in Eq.~\ref{eq:s1} (solid line) fitted to the observed spectroscopic results presented in Fig.~2.
        For the fitted Rydberg fraction of $P_{\ket{S}}=0.16(2)$, we find good agreement using the Rydberg EIT model.
        The data set was recorded for $\Omega_p/2\pi = \SI{168(5)}{\kilo\hertz}$ and $\Omega_c/2\pi = \SI{6.7(6)}{\mega\hertz}$ ($N=80$ repetitions).
        (\textbf{b}) EIT spectroscopy for varying probe Rabi frequencies $\Omega_p/2\pi = 82(3)$, $252(8)$, $385(12)$ and $\SI{1245(40)}{\kilo\hertz}$ with a vertical offset of $0.25$ for each data set added for clarity ($N=4-10$ repetitions).
        The incident probe photons were fixed to $39(2)$ photons/site and $\Omega_c/2\pi = \SI{13.4(6)}{\mega\hertz}$.
        (\textbf{c}) The fitted linewidth of each peak $\Gamma_{\mathrm{EIT}}$ increases for larger $\Omega_p$.
        The black dashed line marks the expected linewidth at $\Gamma_e/2 = 2\pi\times \SI{3.03}{\mega\hertz}$ for the single-particle weak-probe limit.
        (\textbf{d}) The Rydberg fraction $P_{\ket{S}}$ extracted from the fit for increasing $\Omega_p$.
        The black solid line shows the model presented in Eq.~\ref{eq:s2}.
        The horizontal red dashed line in \textbf{d} marks the Rydberg fraction of $P_{\ket{S}} = 0.16$, which is equivalent to the measurement in the main text.
        Error bars denote the standard error of the mean (s.e.m.) for (\textbf{a}, \textbf{b}) and x-error in (\textbf{c},\textbf{d}).
        The y-error bars in (\textbf{c}, \textbf{d}) are the standard deviation (s.d.) of the fit.
    }
    \label{fig:s3}
\end{figure*}
Previous studies have shown that increasing probe powers in EIT involving a Rydberg state result in the breakdown of linear optical response~\cite{Pritchard2010,Peyronel2012}.
This optical nonlinearity arises as Rydberg interactions and the associated energy shifts violate the EIT condition once the energy shift exceeds the width of the EIT transmission feature.
This can be understood as a consequence of the excitation of a delocalized ``Rydberg polariton" with a single photon, which results in an interaction energy shift $U_{\mathrm{int}}(r)$ of the Rydberg transition for subsequent excitations and thus a strong optical nonlinearity at the single-photon level.
This nonlinearity occurs when the Rydberg fraction, which in the single particle limit scales as $\Omega_p^2/(\Omega_p^2 + \Omega_c^2)$, becomes significant.
In Fig.~\ref{fig:s3}a, we compare our data with a simplified description of EIT (dashed line) assuming non-interacting atoms in the weak-probe limit~\cite{Banacloche1995, Fleischhauer2005}.
While this model captures the essential features, there remains a small discrepancy on resonance, which indicates the presence of the Rydberg interaction-induced nonlinearity.
A modified model for EIT in presence of Rydberg interactions was described in~\cite{Petrosyan2011,Garttner2014}.
There, the total susceptibility results from superimposing the standard two-level atom susceptibility
\begin{equation}
\chi^{\mathrm{TA}} = \chi_0 \frac{i\Gamma_e}{\Gamma_e-2i\delta_p}
\end{equation}
with a simplified EIT susceptibility, where the detuning is modified by the Rydberg $S$-state interaction $U^{\mathrm{SS}}_{\mathrm{int}}$
\begin{equation}
\chi^{\mathrm{EIT}} = \chi_0 \frac{i\Gamma_e}{\Gamma_e-2i\delta_p+\Omega_c^2 \left [ \Gamma_{r}-2i(\Delta_2+U^{\mathrm{SS}}_{\mathrm{int}})\right ]^{-1}}.
\label{eq:s0}
\end{equation}
Here, we have defined $\chi_0 = \sigma_0 n_a/k_p$ with $\sigma_0 = 3\lambda_p^2/2\pi$ being the free-space optical cross-section, $n_a$ being the atomic density and $k_p$ the probe wave vector.
$\Gamma_{e,r}$ denotes the natural decay rate of $\ket{e}$ and $\ket{r}$.
Furthermore, $\delta_p = \omega_p-\omega_{eg}$ and $\delta_c = \omega_c-\omega_{re}$ are the single-photon detunings of the probe and control beam and $\Delta_2 = \delta_p+\delta_c$ is the two-photon detuning.
The two contributions are weighted with the Rydberg (ground) state fraction $P_{\ket{S}}$ ($1-P_{\ket{S}}$), which yields the total susceptibility
\begin{equation}
\chi^{\mathrm{REIT}} = P_{\ket{S}}\chi^{\mathrm{TA}} + \left ( 1-P_{\ket{S}} \right ) \chi^{\mathrm{EIT}}.
\label{eq:s1}
 \end{equation}
The lineshapes of the probe response expected from this model are proportional to the imaginary part of the susceptibility, $\mathrm{Im}\left[\chi^{\mathrm{REIT}} \right]$.
Fitting the modified model of Eq.~\ref{eq:s1} with amplitude, offset, $P_{\ket{S}}$, $\Gamma_{\mathrm{EIT}}$, $\Omega_c$ and $U^{\mathrm{SS}}_{\mathrm{int}}$ as free parameters leads to a much better agreement with the measurements (reduced Chi-square of ${\chi_{\mathrm{red}}^2 \approx 1.5-1.7}$) compared to the EIT model neglecting Rydberg interactions (${\chi_{\mathrm{red}}^2 > 3}$), see Fig.~\ref{fig:s3}a.
As a  result from the fit, we obtained ${\Gamma_{\mathrm{EIT}} = \Gamma^{\mathrm{fit}}_e/2 = 2\pi\times 2.97(16)\,\si{\mega\hertz}}$ (see main text),  $\Omega_c/2\pi = \SI{6.7(6)}{\mega\hertz}$ and $P_{\ket{S}} = 0.16(2)$.
To probe the nonlinearity in detail, we investigated the dependence of the EIT spectra on the probe Rabi frequency $\Omega_p$ at a fixed control Rabi frequency of ${\Omega_c/2\pi = \SI{13.4(6)}{\mega\hertz}}$, see Fig.~\ref{fig:s3}b.
For increasing $\Omega_p$, we observe a broadening of the absorption feature, consistent with previous studies in the same regime (${\Omega_c \approx \Gamma_e}$)~\cite{DeSalvo2016,Tebben2021}, and the disappearance of the transmission window at ${\Omega_p/2\pi = \SI{1.24(4)}{\mega\hertz}}$.
The fitted linewidth $\Gamma_{\mathrm{EIT}}$ broadens rapidly with increasing $\Omega_p$ and equals ${\Gamma_{\mathrm{EIT}}/2\pi = \SI{6.9(15)}{\mega\hertz}}$ at the breakdown point where the EIT double peaks vanish, see Fig.~\ref{fig:s3}c.
Next to the linewidth, we also extract the Rydberg fraction $P_{\ket{S}}$ via the fit and observe an increase of $P_{\ket{S}}$ with increasing $\Omega_p$, illustrated in Fig.~\ref{fig:s3}d.\\
We compare $P_{\ket{S}}$ with a model including collective enhancement~\cite{Petrosyan2011}, which predicts
\begin{equation}
P_{\ket{S}} = \frac{n_{SA}\Omega_p^2\Omega_c^2}{n_{SA}\Omega_p^2\Omega_c^2+[\Omega_c^2-4\delta_p\Delta_2]^2+16\Delta_2^2\gamma_e^2},
\label{eq:s2}
\end{equation}
where $n_{SA}=(1/a^2_{\mathrm{lat}})\times \pi(r^{\mathrm{SS}}_b)^2 \approx 77$ denotes the atom number in the blockade volume, $r^{\mathrm{SS}}_b= \sqrt[6]{2C^{\mathrm{SS}}_6\Gamma_e/\Omega_c^2}$ is the EIT-blockade radius and $C^{\mathrm{SS}}_6 = \SI{3.15}{\giga\hertz\micro\meter\tothe{6}}$ the van der Waals coefficient between two atoms in state $\ket{S}$.
The fraction $P_{\ket{S}}$ predicted by the model is below the results obtained from the fit.
We attribute this discrepancy to two main sources:
First, the presented model is well suited to describe the data sets for $\Omega_p/2\pi \leq \SI{385}{\kilo\hertz}$, where the reduced Chi-square amounts to $\chi_{\mathrm{red}}^2 \approx 1.5-1.9$, whereas for the largest $\Omega_p$ we find $\chi_{\mathrm{red}}^2 \approx 4.8$.
We conclude that for larger $\Omega_p$ the model requires further assumptions to capture the essential physics.
We note however, that the presented measurements in the main text are performed for $\Omega_p$ well below the breakdown of EIT, where we find the model to hold and where we observe minor Rydberg fractions of $P_{\ket{S}} = 0.16(2)$, indicated by the horizontal red dashed line in Fig.~\ref{fig:s3}d.
Second, the measurements were performed at a lattice depth of $300\,E_r$ with a relatively large incident photon flux of $39(2)$ photons/site.
%
\begin{figure*}[t!]
    \centering
    \includegraphics{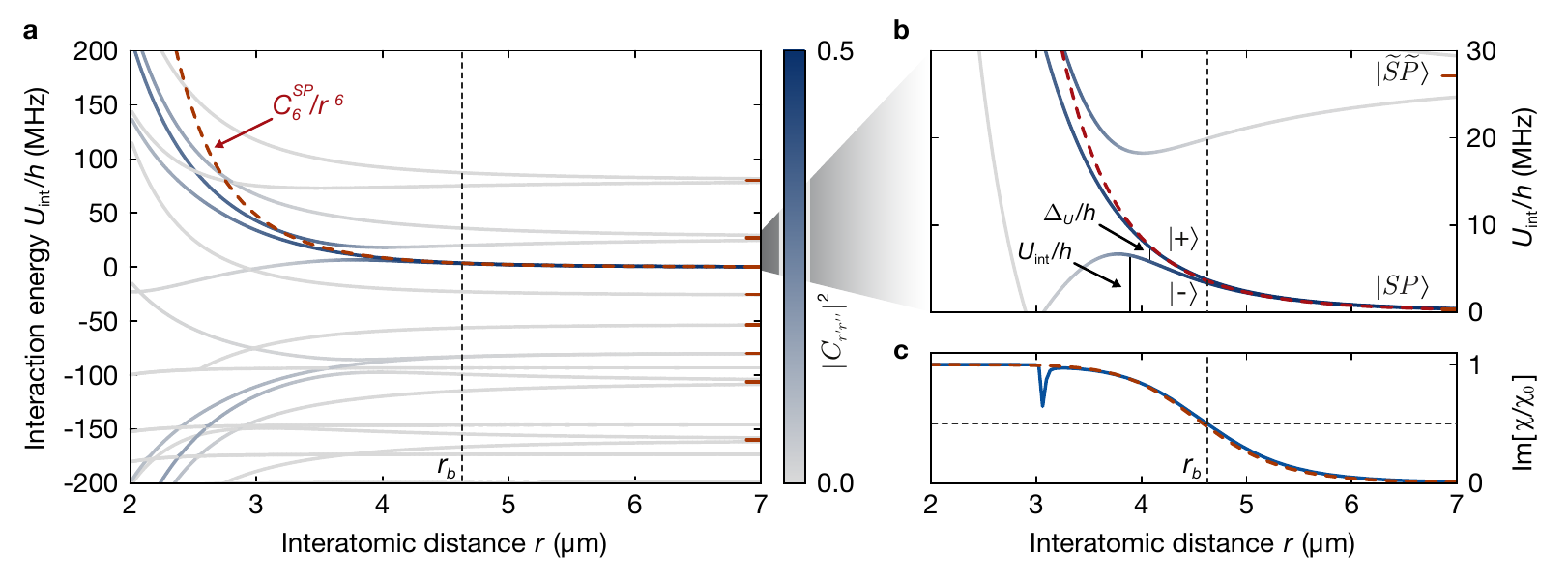}
    \caption{\textbf{Calculated interaction potentials and optical response function.}
        (\textbf{a})  Potential energy curves relevant for coupling the $44S_{1/2}$ to the $44P_{3/2}$ Rydberg state in an external magnetic field of $B_z = 28.5\, G$. 
        The intensity of the blue coloring indicates the relative optical coupling strength $\left | C_{r'r''} \right |^2$ for the spin states $\ket{S}$ and $\ket{P}$.
        The red ticks at a large distance mark the magnetic sublevel splittings.
        The red dashed curve resembles a van der Waals potential $C_6^{\mathrm{SP}}/r^6$ with $C^{\mathrm{SP}}_6/h \approx \SI{35}{\giga\hertz\  \micro\meter^6}$.
        (\textbf{b}) Potential energy curves in the close vicinity of the EIT-blockade radius $r_b$ (vertically-dashed line). The red ticks mark the asymptotic state $\ket{SP}$ and $\ket{\tilde{r}_S\tilde{r}_P}$, where $\ket{\tilde{r}_S} = \ket{44S_{1/2}, m_J=+1/2}$ and $\ket{\tilde{r}_P} = \ket{44P_{3/2}, m_J=+1/2}$.
        The splitting between the curves marked as $\ket{+}$ and $\ket{-}$ amounts to $\Delta_U/h$ and gives rise to the dipolar $\ket{S}-\ket{P}$ exchange.
        (\textbf{c}) Imaginary part of the probe susceptibility ($\ket{g} \leftrightarrow \ket{e}$) for variable Rydberg-Rydberg distance and $\delta_p = 0$.
        The blue curve takes the four most relevant Rydberg pair potentials with $\left | C_{r'r''} \right |^2>0.05$ into account, resulting in an EIT-blockade of $r_b = \SI{4.63}{\micro\meter}$ (vertically-dashed line) where $\mathrm{Im}[\chi/\chi_0](r_b) = 1/2$.
        Approximating the potentials by an effective van der Waals potential with $C^{\mathrm{SP}}_6/h \approx \SI{35}{\giga\hertz\  \micro\meter^6}$, we find excellent agreement of the optical response as shown by red dashed curve.
        Note that the dipole-dipole interaction coefficient $C_3$ becomes negligible for $r>\SI{4}{\micro\meter}$, as resonant dipole-dipole coupling between $\ket{S}$ and $\ket{P}$ is forbidden as the difference in magnetic quantum number amounts to $\Delta m_J = 2$.
    }
    \label{fig:s4}
\end{figure*}
Here, we observed a decrease of the atomic filling to $\approx 0.7$, which can influence the transmission detection.
The data in the main text was measured at much lower probe photon of $0.36(2)$ photons/site where such heating was negligible, resulting in a smaller linewidth and transmittance dip.

\subsection{Interaction potentials and derivation of the EIT-blockade radius}
We calculated the interaction potentials for Rydberg pair states $\ket{SP}$ and $\ket{PS}$ using the open-source program ``Pairinteraction"~\cite{Weber2017}. 
The software performs exact diagonalization of the electrostatic interaction Hamiltonian between two Rydberg atoms.
The interactions are calculated beyond the Leroy radius, meaning the model assumes two independent charge distributions which can be expressed by a multipole expansion.
To calculate the Rydberg-Rydberg interaction for a given internuclear distance, the program creates and diagonalizes larger matrices where numerous unperturbed states $\ket{n_1l_1j_1m_{j1};n_2l_2j_2m_{j2}}$ are coupled with the target state $\ket{SP}$ according to the selection rules.
Within the diagonalized basis, one can then further calculate the overlap with $\ket{SP}$ which is equivalent to the optical coupling strength, given by the blue coloring in Fig.~\ref{fig:s4}.\\
For Rydberg pair states composed of opposite parity Rydberg states, interactions are typically described by resonant dipole-dipole interactions following a $\propto1/r^3$\,-scaling.
In our case however, the first-order dipole-dipole matrix elements vanish as the magnetic quantum numbers of $\ket{P}$ and $\ket{S}$ differ by $\Delta m_J = 2$.
As a consequence, the interactions are described in second order perturbation theory by van der Waals interactions $C^{\mathrm{SP}}_6/r^6$, originating from off-resonant coupling to pair states formed by neighboring Zeeman sublevels.
Due to the constant bias field of $B_z = 28.5\,$G, these magnetic Zeeman sublevels are split by tens of Megahertz, see Fig.~\ref{fig:s4}a and b.
Fitting the resulting interaction potential for $r > \SI{4}{\micro\meter}$, we find a van der Waals coefficient of $C^{\mathrm{SP}}_6/h \approx \SI{35}{\giga\hertz\  \micro\meter^6}$, see red dashed line in Fig.~\ref{fig:s4}a and b.\\
Using the calculated interaction potentials, we derive the EIT-blockade radius $r_b$ by analysing the single-particle optical response function $\mathrm{Im}[\chi]$ of the probe transition.
We model our system with all relevant Rydberg pair states, assuming $\delta_{p} = \delta_{c} = 0$ and Rydberg interactions to neighboring pair states as $U_{\mathrm{int}}^{r'r''}(r)$ with the Hamiltonian
\begin{equation}
\begin{split}
H =& \frac{\hbar}{2} \Omega_p\ket{e}\bra{g} + \frac{\hbar}{2}\sum_{r',r''} \Omega_c C_{r'r''} \ket{r'r''}\bra{e}\\
&+ \frac{\hbar}{2\pi}\sum_{r',r''}U_{\mathrm{int}}^{r'r''}(r)\ket{r'r''}\bra{r'r''} +h.c..
\end{split}
\end{equation}
Here, $\ket{r'r''}$ are Rydberg pair states and ${C_{r'r''} = \braket{r'r''}{SP}}$ is the overlap of $\ket{r'r''}$ with the bare state $\ket{SP}$.
By accounting for Rydberg pair states $\ket{r'r''}$ with $\left | C_{r'r''} \right |^2>0.05$ and $U_{\mathrm{int}}^{r'r''}<\SI{100}{\mega\hertz}$, we calculated the steady-state solution of the probe transition density matrix ($\rho_{eg}$) using the QuTiP package~\cite{Johansson2013} and obtained the distance dependence of the optical probe response $\mathrm{Im}[\chi](r)$ with $\chi = -2\chi_0\gamma_e\rho_{eg}/\Omega_p$, see Fig.~\ref{fig:s4}c.
Defining the EIT-blockade radius $r_b$ at the point where $\mathrm{Im}[\chi/\chi_0]$ reduces to half of its maximum value, we find an EIT-blockade radius of $r_b = \SI{4.63}{\micro\meter}$, in agreement with $r_b = (2C_6\Gamma_e/\Omega_c^2)^{1/6} = \SI{4.6}{\micro\meter}$ for $C^{\mathrm{SP}}_6/h \approx \SI{35}{\giga\hertz\  \micro\meter^6}$.
This estimated blockade radius is also in good agreement with the experimental measurements in Fig.~1c and Fig.~5.
We note that the simplified model of pure van der Waals interactions $C^{\mathrm{SP}}_6/r^6$, with $C^{\mathrm{SP}}_6 \approx \SI{35}{\giga\hertz\  \micro\meter^6}$, leads to a very similar spatial optical response, see the red dashed line in Fig.~\ref{fig:s4}c.

%
%
\subsection{Estimating the $\ket{S}-\ket{P}$ exchange rate}
As described in the main text, for larger atomic arrays, the measured radial transmission profile in Fig.~5b deviated from the theoretical expectation.
In those systems, we observed an overall reduction of the central transmission as compared to the small array, see Fig.~5a.
In addition, we observed a ``halo" of reduced transmission reaching to distances beyond the ancilla-induced EIT-blockade radius $r_b$.
We attribute these observations to dipolar exchange processes between the ancilla atom in state $\ket{P}$ and the surrounding atoms dressed to $\ket{S}$~\cite{Gunter2013,Schempp2015,Schonleber2015}.
In contrast to previous studies, where the exchange process was given by resonant dipole-dipole coupling, in our case the exchange arises only in second-order as the magnetic Zeeman sublevels of the Rydberg state $\ket{S}$ admixed to the mirror and the ancilla prepared in $\ket{P}$ differ by $\Delta m_J = 2$.
In the following, we quantitatively estimate the effective exchange rate $J^{\mathrm{eff}}_{\mathrm{ex}}$ mediated by off-resonant coupling to other Zeeman sublevels.\\
In a first step, we consider the dynamics of two isolated Rydberg atoms in states $\ket{S}$ and $\ket{P}$ and derive the exchange coupling rate $J_{\mathrm{ex}}$.
For distances ${r>\SI{4}{\micro\meter}}$, only two pair-potentials are relevant, see Fig.~\ref{fig:s4}b.
These potentials can be decomposed in the symmetric ${\ket{+} = \frac{1}{\sqrt{2}} \left ( \ket{SP}+\ket{PS} \right )}$ and anti-symmetric ${\ket{-} = \frac{1}{\sqrt{2}} \left ( \ket{SP}-\ket{PS} \right )}$ superposition of the $\ket{SP}$ pair basis.
Initializing the atoms in the bare state $\ket{SP} = \frac{1}{\sqrt{2}}(\ket{+}+\ket{-})$ results in a coherent exchange to a state $\ket{PS}$ with rate $J_{\mathrm{ex}}/(2\pi) = \Delta_{U}/(2h)$, where $\Delta_{U}$ is the energy splitting of the two pair-potentials.
%
\begin{figure}[t!]
    \centering
    \includegraphics{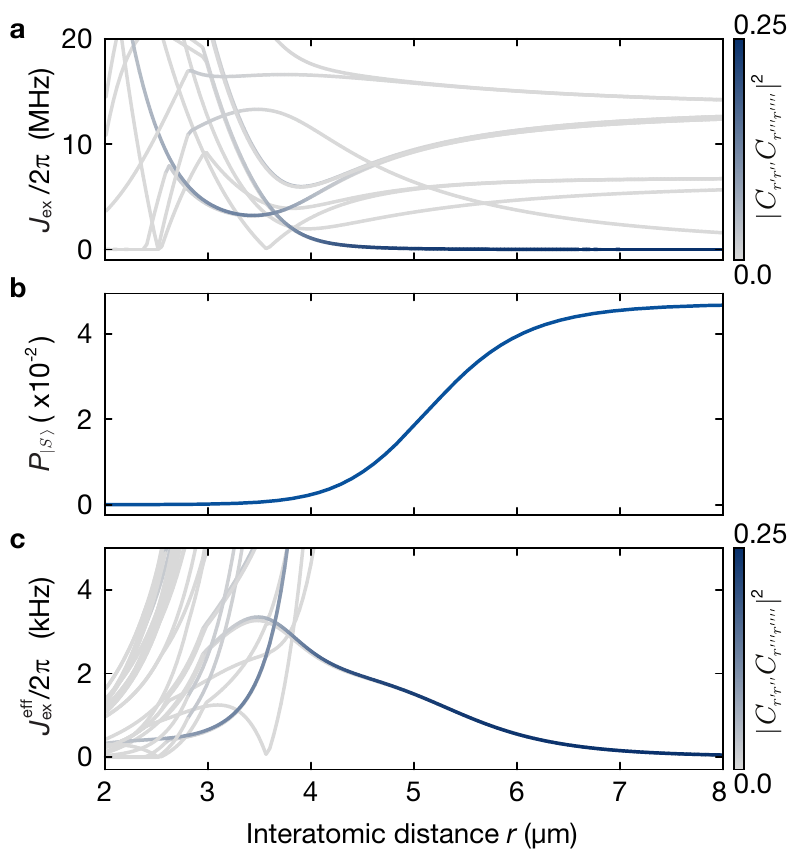}
    \caption{\textbf{Dipolar exchange.}
     (\textbf{a}) $\ket{S}-\ket{P}$ exchange rate $J_{\mathrm{ex}}$.
     The rate is calculated from the splitting between neighboring interaction potentials ($U_{\mathrm{int}}^{r'r''}$ and $U_{\mathrm{int}}^{r'''r''''}$), while the intensity of the blue coloring indicates the overlap $\left | C_{r'r''} C_{r'''r''''} \right |^2$ with the bare state $\ket{SP}$.
     (\textbf{b}) $\ket{S}$ Rydberg fraction as calculated by Eq.~\ref{eq:s2}.
     (\textbf{c}) Effective exchange rate, given by the product of the upper and middle graph $J^{\mathrm{eff}}_{\mathrm{ex}}(r) = P_{\ket{S}}(r)\times J_{\mathrm{ex}}(r)$.
     For larger distances $r>\SI{4}{\micro\meter}$, the effective exchange rate is predominantly given by the difference of the symmetric and antisymmetric pair states $\ket{+}$ and $\ket{-}$, respectively.
     For smaller distances $r<\SI{4}{\micro\meter}$ however, a multitude of pair states contribute, resulting in strong dephasing.
    }
    \label{fig:s5}
\end{figure}
The coherent exchange dynamics of the bare states is thus described by
\begin{equation}
\ket{\Psi}(t) \approx \mathrm{cos}(J_{\mathrm{ex}}t)\,\ket{SP}+e^{i\phi}\mathrm{sin}(J_{\mathrm{ex}}t)\,\ket{PS},
\label{eq:sx}
\end{equation}
where $\phi$ is a global phase. 
For smaller distances (${r\leq \SI{4}{\micro\meter}}$), a multitude of pair states contribute to $J_{\mathrm{ex}}$.
In principle, any two pair potentials $U_{\mathrm{int}}^{r'r''}$ and $U_{\mathrm{int}}^{r'''r''''}$ with energy splitting $\Delta_{U}^{r'r''r'''r''''} = | U_{\mathrm{int}}^{r'r''} - U_{\mathrm{int}}^{r'''r''''} |$ will result in a coherent $\ket{S}-\ket{P}$ exchange, as long as the overlaps $C_{r'r''}$ and $C_{r'''r''''}$ are significantly large.
Focusing only on pair states with $\left | C_{r'r''} \right |^2>0.01$, we find the exchange rates as illustrated in Fig.~\ref{fig:s5}a.
While for larger distances the exchange is restricted to two pair states ($ \ket{SP}\leftrightarrow \ket{PS}$), a multitude of states are contributing for shorter distances resulting in strong dephasing and possible coupling to other states, such as $\ket{SP} \rightarrow \ket{r'r''}$. \\
In a next step, we estimate the Rydberg fraction $P_{\ket{S}}(r)$ quantifying the probability to find an atom surrounding the ancilla atom in Rydberg state $\ket{S}$ using Eq.~\ref{eq:s2}.
Due to the large interaction energy shift between atoms in Rydberg states $\ket{S}$ and $\ket{P}$, $P_{\ket{S}}(r)$ vanishes for small distances, see Fig.~\ref{fig:s5}b.
Finally, we combine the exchange rate $J_{\mathrm{ex}}(r)$ with the distance-dependent probability $P_{\ket{S}}(r)$ to an effective dipolar exchange rate by $J^{\mathrm{eff}}_{\mathrm{ex}}(r) = P_{\ket{S}}(r) \times J_{\mathrm{ex}}(r)$~\cite{Gunter2013}, resulting in a maximum rate of $J^{\mathrm{eff}}_{\mathrm{ex}}/2\pi \approx \SI{3.2}{\kilo\hertz}$ peaked at around $r \approx \SI{3.7}{\micro\meter}$ (see Fig.~\ref{fig:s5}c).\\
\begin{figure*}[t!]
    \centering
    \includegraphics{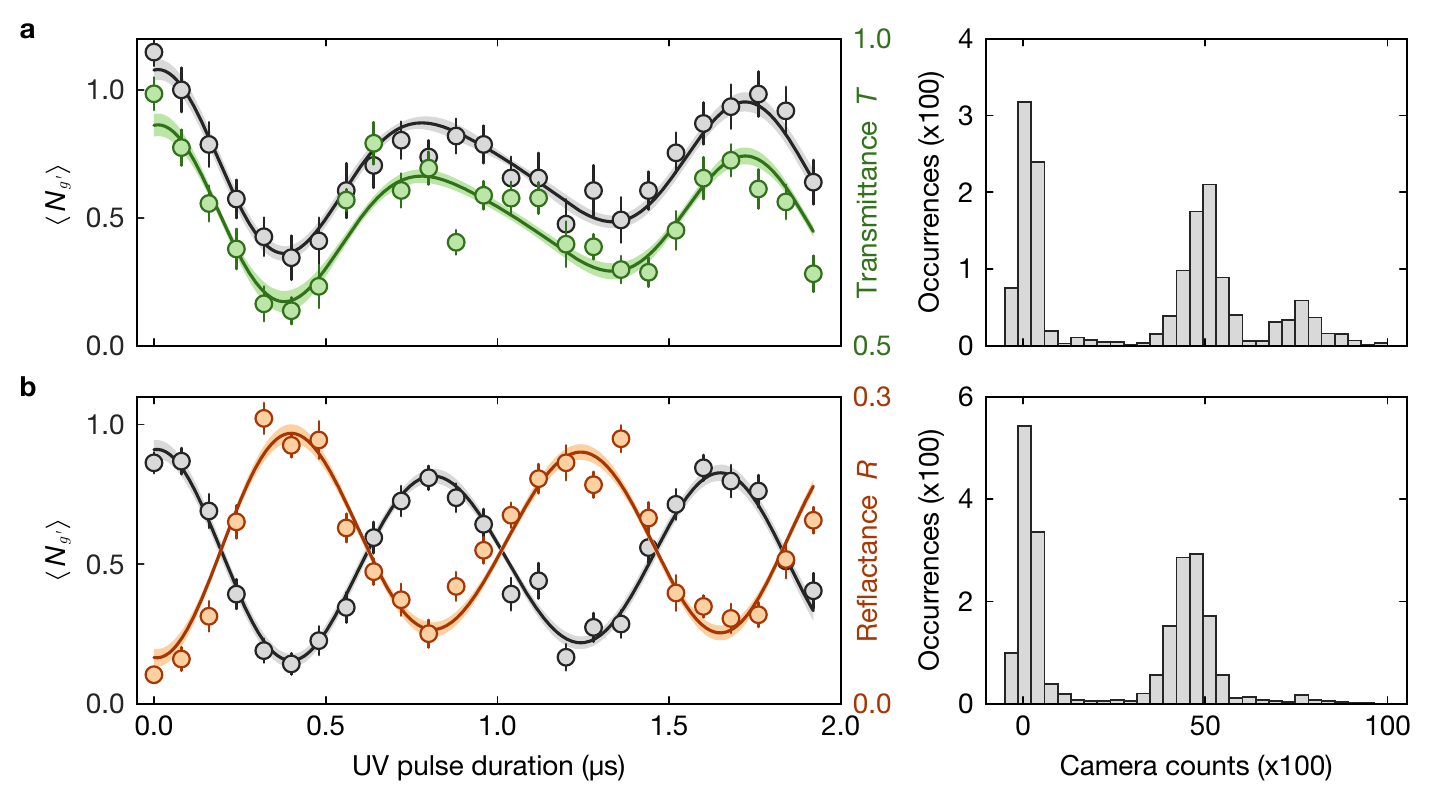}
    \caption{\textbf{Rabi oscillation and fluorescence signal count histogram.}
        (\textbf{a, left}) Detected number of atoms in $\ket{g'}$ during ancilla Rabi oscillation (gray) and corresponding transmittance oscillation (green)($N=60$ repetitions).
        (\textbf{b, left}) Detected number of atoms in $\ket{g'}$ during ancilla Rabi oscillation (gray) and corresponding reflectance oscillation (brown) ($N=80$ repetitions).
        (\textbf{Right}) Camera count histograms of fluorescence images evaluated from ancilla Rabi oscillation data shown on the left.
        The histograms provide the cutoff of the atom number used in the Rabi oscillation analysis.
        The transmission data set shows a probability of $0.28(5)$ for having initially two ancilla atoms resulting in a beating frequency due to collectively enhanced oscillations. The dynamics is in excellent agreement with the transmittance oscillation.
        On the contrary, the reflection data set has improved preparation fidelities with negligible two atom fraction and consequently exhibits a pure single-frequency sinusoidal Rabi oscillation.
        Error bars denote the s.e.m..
    }
    \label{fig:s6}
\end{figure*}
To compare to the dynamics in our array, we extend this two-particle model towards a larger many-body system, where the ancilla can undergo coherent exchange $J^{\mathrm{eff}}_{\mathrm{ex}}(r)$ with all atoms in the vicinity.
We include an incoherent coupling $\gamma(r)$ to the environment and solve the resulting Lindblad master equation using the QuTiP package~\cite{Johansson2013}.
We find the dynamics of the ancilla to be well described by an effective exchange rate of $J_{\mathrm{ex}}^{\mathrm{col}} \approx 2\pi\times \SI{30}{\kilo\hertz}$.
In contrast to two-atom exchange, here the dynamics is collectively enhanced as a consequence of the indistinguishability of all coupled atoms at a certain distance from the ancilla.
A single exchange is concluded after a duration of $ 1/(2\times \SI{30}{\kilo\hertz})\approx \SI{16}{\micro\second}$, which is on the order of our probe duration ($t_p = \SI{20}{\micro\second}$).
As the ancilla undergoes collective exchange with all atoms at around $r\approx \SI{3.7}{\micro\meter}$, the original $\ket{P}$ excitation at the center of the array quickly delocalizes onto a ring.
Starting from there, the excitation then again undergoes collective exchange, ultimately leading to a complete loss of $\ket{P}$-state fraction at the center.
To suppress such exchange processes, one can either operate on shorter probe timescales or reduce the probe power, as the exchange process scales with $\propto \Omega_p^2$ in the limit $\Omega_p\ll\Omega_c$~\cite{Schempp2015}.

\section{Data evaluation}
In this section, we provide general information on how the data was analysed and also describe simulation methods used in the main text.\\
For the transmittance and reflectance analysis, we used a region-of-interest (ROI) of $5\times5$ binned pixels, equivalent to an area of $5\times\SI{5}{\micro\meter}$ that is centered around the ancilla atom.
The ROI area covers $\approx 100$ lattice sites, smaller than the atomic array, which has a diameter of $\SI{9.4(14)}{\micro\meter}$.
For each measurement, we recorded and subtracted a background image, which removed an offset on the camera signal. 
We then evaluated the transmittance and reflectance by comparing it with a reference image including the probe beam, yet after dropping the atoms.
%
\subsection{Details on the observed Rabi oscillations}
The single-atom Rabi oscillation was detected indirectly through the loss of the Rydberg-excited ancilla atom due to the anti-trapping of the Rydberg state.
After the UV pulse and the measurement of the transmittance or reflectance, we removed the surrounding array atoms in the ground state $\ket{g}$ by a resonant push-out pulse and waited several milliseconds to ensure that the Rydberg state $\ket{P}$ left the trap.
Subsequently, we detected the atoms in $\ket{g'}$, the only state which had not been removed, using site-resolved fluorescence imaging.\\
Investigating the oscillation of the transmission signal presented in Fig.~3 reveals a slight discrepancy from the presented single-frequency sinusoidal model.
To obtain further insight, we analyse the histogram of fluorescence counts in a region of $3\times 3$ lattice sites around the ancilla atom, see Fig.~\ref{fig:s6}.
Interestingly, for transmission, the data set shows three separate peaks.
We identify the two left peaks as $N_{g'}=0$ and $N_{g'}=1$ atoms in $\ket{g'}$.
The rightmost peak is located between the signal corresponding to $N_{g'}=1$ and the expected value for $N_{g'}=2$ atoms.
This enlarged signal can be explained by the presence of two atoms occupying different adjacent sites in the vertical lattice, originating from imperfect preparation of the initial state.
We verified this by intentionally preparing systems of atoms occupying neighboring layers of the vertical lattice, which results in identical distributions.
Using thresholds to discriminate between $N_{g'}=0$, $N_{g'}=1$ and $N_{g'}=2$ based on the fluorescence count histograms presented in Fig.~\ref{fig:s6}, we obtain probabilities of $0.42$, $0.44$ and $0.14$, for the three cases averaged over the time evolution, respectively.\\
In our standard reconstruction of lattice site occupancy, the two cases $N_{g'}=1$ and $N_{g'}=2$ would not be discriminated.
We therefore reevaluated the atom numbers contributing to the oscillating $\langle N_{g'}(t)\rangle$ with the new thresholds for $N_{g'}$, see Fig.~\ref{fig:s6}a.
First, this procedure introduces a contribution with $N_{g'}>1$ resulting in an overall positive offset to the Rabi oscillation signal.
Second, the Rabi oscillation signal $\langle N_{g'}(t)\rangle$ exhibits a deviation from a pure sinusoidal single-frequency oscillation, in much better agreement with the observed time-dependent transmission data $T(t)$. \\
Based on this observation, we also extend our model to describe the observed transmissions $T(t)$ presented in Fig.~3.
We allow for a beating of the Rabi oscillation between the initial configurations $N_{g'}=1$ and $N_{g'} = 2$ originating from the $\sqrt{N_{g'}}$ enhancement of the Rabi frequency~\cite{Gaetan2009,Zeiher2015}.
Their relative amplitudes, which are directly connected to the probability of $N_{g'}=0$, $N_{g'}=1$ and $N_{g'} = 2$, were left as free fit parameters.
From the fit, we obtain an initial fraction of $P_2(t=0) = 0.28(5)$ for $N_{g'} = 2$.
During the dynamics, this initial fraction averages to $\langle P_2(t) \rangle = 0.14(3)$, in excellent agreement with the value obtained from the histogram.
For the reflection data set, we improved the initial state preparation to suppress any $N_{g'}=2$ occupation, resulting in pure single frequency oscillations.\\
%

%
\subsection{Simulating the photon number histogram}
In Fig.~4 we recorded two distinct histograms of the detected photon numbers in reflection, with either the ancilla being prepared in the ground or the Rydberg state.
In this section, we derive and describe the Monte-Carlo simulation used to estimate the shape of the recorded histograms.
In these measurements, we chose a prolonged interaction time of $t_p = \SI{60}{\micro\second}$.
%
%
With the ancilla prepared in $\ket{g'}$, the system is transparent and all collected photons within the ROI can be attributed to imperfections in the EIT-signal, such as self-blockade, see the blue histogram in Fig.~4.
The probability distribution follows a Poissonian distribution with a variance exceeding photon shot noise by a factor of two due to the stochastic EMCCD amplification process~\cite{Robbins2003}.
We find a signal of $\mu = 18.3(7)$ photons, which is slightly larger than the photon background count of $16.0(7)$ that was globally subtracted in Fig.~4.\\
For the ancilla in $\ket{P}$, the recorded reflection values are shifted towards higher values because the mirror is switched reflective (orange histogram in Fig.~4).
We now observe a broad asymmetric distribution featuring a maximum at low photon numbers and an additional tail towards larger photon numbers.
To reproduce the detected shape, we repeatedly run the following steps to generate a simulated photon number histogram.
As a starting point, we assume the photon numbers to be a combination of background photons and photons resulting from reflection, picking two values from two distinct Poissonians and sum them with their respective weights.
The first Poissonian has a mean of $\mu = 18.3(7)$ photons, whereas the mean of the second ``switched" Poissonian remains to be determined.
The relative weight between these two contributions is given by the preparation fidelity of the ancilla in $\ket{P}$ and the Rydberg decay time.
We obtain these values experimentally by performing independent reference measurements (see inset of Fig.~4), resulting in an ancilla $\ket{P}$ preparation fidelity of $0.85(10)$ and a Rydberg lifetime of $\tau = \SI{27(6)}{\micro\second}$.
These experimental numbers directly enter our simulation, by means of randomly sampling the initial Rydberg fraction as well as the moment of the Rydberg decay within their respective error bars for each realization from a Gaussian distribution.\\
Running the aforementioned procedure for $10^5$ independent realizations, we can first estimate the mean photon number for the switched mirror.
We do so by varying the mean photon number of the second Poissonian until the simulation best fits larger photon numbers exceeding $>60$.
Here we find best agreement based on a maximum likelihood estimate if the second ``switched" Poissonian has a mean value of $65(2)$ photons.
The resulting distribution in absence of Rydberg decay and with perfect ancilla preparation is shown in an orange dashed curve in Fig.~4. \\
Finally, we derive the theoretical shape of the histogram by performing $5000$ independent runs, each containing $10^4$ independent sets of photon numbers.
Having these $5000$ realizations, we can then derive a mean value and standard deviation resulting in the orange solid curve and shaded region in Fig.~4.

\subsection{Rydberg lifetime of Rydberg $\ket{P}$-state}
We measured the Rydberg lifetime of the $\ket{P}$-state by varying the time interval $\delta t$ between the ancilla excitation pulse and the detected transmission of the probe.
We observed an increase of the transmittance due to the finite Rydberg lifetime, which continuously reduces the Rydberg fraction $P_{\ket{P}}$, see Fig.~4 inset.
We estimated the Rydberg fraction $P_{\ket{P}}$ via
\begin{equation}
P_{\ket{P}}(\delta t) = \frac{\eta_{\mathrm{init}}}{t_p} \int_{\delta t}^{\delta t + t_p} e^{-t'/\tau}dt'.
\end{equation}
Here, $\eta_{\mathrm{init}}$ is the efficiency of initially preparing the ancilla atom in $\ket{P}$.
We fit the transmittance, which is proportional to the Rydberg fraction $P_{\ket{P}}(\delta t)$, with the initial preparation efficiency $\eta_{\mathrm{init}}$, the Rydberg decay $\tau$ and an offset as free fit parameters.
The transmittance is constrained in the fit to lie between the transmittance of the cooperative mirror and the EIT feature on resonance. 
From the fit, we obtain an initial preparation efficiency of $\eta_{\mathrm{init}} = 0.85(10)$ and the Rydberg decay of $\tau = \SI{27(5)}{\micro\second}$.
For these values and $\delta t = \SI{4}{\micro\second}$, we reproduce the experimental setting of Fig.~2b, resulting in $P_{\ket{P}}=0.52(8)$.
The fitted $\eta_{\mathrm{init}}$ is in agreement with the independent measurement of \ket{g'} preparation efficiency of $0.83(4)$, assuming perfect excitation from $\ket{g'}$ to $\ket{P}$
Comparing the Rydberg lifetime $\tau$ of $\ket{P}$ with a theoretical estimation of $\SI{65}{\micro\second}$ at $T = 300\,K$ from ``ARC" package~\cite{Sibalic2017} and an experimental measurement of $\SI{64.2(26)}{\micro\second}$ of $^{85}\mathrm{Rb}$ atoms in a magneto-optical trap~\cite{Branden2009}, our measured lifetime is lower by approximately a factor of two.\\
We estimated that two possible factors, photoionization and motion of the ancilla atom, can contribute to this discrepancy.
We investigated photoionization of a Rydberg state by the optical lattices via a loss rate measurement in a dilute atomic cloud by off-resonant dressing to the Rydberg state, $\delta_{\mathrm{UV}} \gg \Omega_{\mathrm{UV}}$, for variable lattice depths up to $1000\,E_r$.
Our observation of similar loss rates independent of the lattice depths underlines the inefficiency of the Rydberg photoionization by the optical lattices.
Furthermore, we explored the motional dynamics of the Rydberg ancilla atom in the 1D optical lattice with a depth of $100\,E_r$ by numerically solving the time evolution of the initial wave packet.
Due to the anti-trapping of the Rydberg state, the initial wave packet in a single lattice site at $t=\SI{0}{\micro\second}$ expands after excitation up to $\Delta d \approx \SI{5.3}{\micro\meter}$ at $t=\SI{120}{\micro\second}$.
The expansion of the ancilla shifts the blockade area out of the ROI which results in a slightly increased measured transmittance. The in-plane expansion has a larger influence compared with the vertical expansion due to the momentum kick of the UV photon.
However, the ancilla motion alone could not fully explain the cause of shortened lifetime.
As a consequence, further possible effects need to be investigated in the future.
\begin{figure}[t!]
    \centering
    \includegraphics{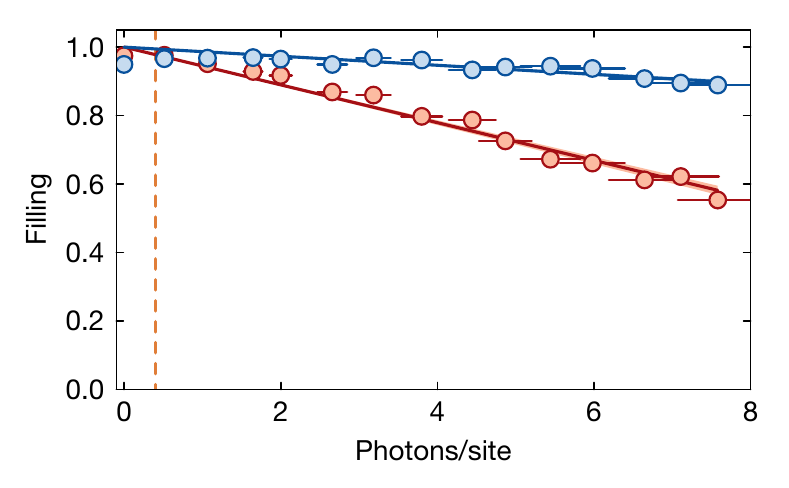}
    \caption{\textbf{Atom loss during mirror operation.}
        Average atomic filling after illuminating the sample with a variable amount of resonant probe photons for either the cooperative mirror (red) or EIT configuration (blue) where an additional $\SI{480}{\nano\meter}$ beam is applied.
        Applying a linear fit, we obtain a loss rate of $0.055(2)$ atoms/photon and $0.013(1)$ atoms/photon for the cooperative mirror and the EIT, respectively.
        The vertical orange line marks $0.36(2)$ photons/site for a duration of $t_p = \SI{20}{\micro\second}$, which was used for all transmission measurements except the photon number histogram.
        The measurements are averages over $N = 23$ independent repetitions.
        Error bars denote the s.e.m..
    }
    \label{fig:s7}
\end{figure}

\section{Additional measurements}
Here we show additional measurements to provide more details about the heating of atoms in the array due to probe photons, the reflectance of disordered atoms via Bloch oscillations, and derive the camera signal to photon conversion factor.
\subsection{Atom loss due to heating by the probe photons}
To quantify potential probe-beam induced atom losses, we illuminated the atomic array at a lattice depth of $40\,E_r$ for a variable amount of resonant probe photons, see Fig.~\ref{fig:s7}.
For the case of the cooperative mirror, where only the probe beam was on, the loss rate amounted to $0.055(2)$ atoms/photon.
In contrast, the EIT configuration had a lower loss rate of $0.013(1)$ atoms/photon as expected from the transparency window.
Since the switched mirror case is a statistical combination of these two cases, the atom loss is estimated to be within the bounds of these two cases depending on the ancilla Rydberg fraction $P_{\ket{P}}$.
The vertical orange line marks $0.36(2)$ photons/site corresponding to $\Omega_p/2\pi = \SI{168(5)}{\kilo\hertz}$ for a duration of $t_p = \SI{20}{\micro\second}$, used for most measurements in the main text at a lattice depth of $100\,E_r$.
For these parameters, we observe negligible atomic loss, thus allowing in principle for multiple detections before the reduction in atomic density significantly deteriorates the subwavelength mirror performance.


\subsection{Inducing spatial disorder through Bloch oscillations}
\begin{figure}[t!]
    \centering
    \includegraphics{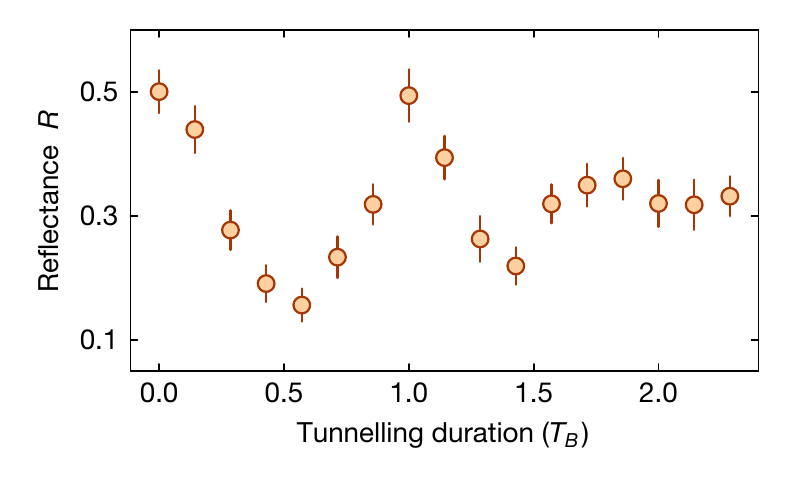}
    \caption{\textbf{Reflectance under vertical Bloch oscillation.}
    The Bloch oscillation at the half period ($0.5T_B$) breaks the cooperative response by maximally displacing the atoms vertically, resulting in isotropic scattering of the disordered atoms.
    The resonant reflectance at half period amounts to $0.16(3)$.
    The measurements are averages over $N = 13$ independent repetitions.
    Error bars denote the s.e.m..
    }
    \label{fig:s8}
\end{figure}
To compare the cooperative response with the dissipative free-space scattering from uncorrelated, disordered atoms, we performed Bloch oscillations along the propagation direction of the probe beam to allow vertical position spread.
We started with the 2D ordered array at a depth of $15\,E_r$ and $20\,E_r$ in the vertical and horizontal lattices, respectively.
We instantaneously reduced the vertical depth to $4\,E_r$ letting the atoms dynamically oscillate under the potential energy difference between adjacent lattice sites of $\Delta_z/h = ~\SI{360}{\hertz}$ along the vertical axis, arising from the magnetic field and gravity gradient.
The ordered array spread out vertically and refocused after the Bloch period of $T_B = h/\Delta_z \sim\SI{2.8}{\milli\second}$ with an estimated maximum half width of $d_z = 4Ja_{\mathrm{lat}}/\Delta_z \sim\SI{3.6}{a_{\mathrm{lat}}}$, where $J$ is the tunnelling rate in the vertical direction and $a_{\mathrm{lat}}$ is the lattice constant.
At half period ($0.5 T_B$), the spread of the atoms along the vertical direction is maximal, resulting in a reflectance of $0.16(3)$, being equivalent to the expected isotropic scattering of $\approx 0.13$ for a single particle and the NA of the objective of $0.68$.

\subsection{Camera signal to photon conversion}
\begin{figure}[t!]
    \centering
    \includegraphics{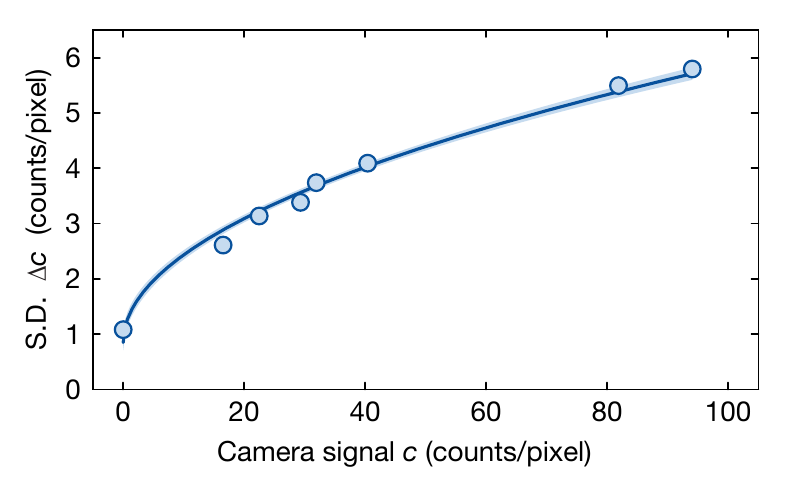}
    \caption{\textbf{Calibration of camera conversion.}
        Standard deviation (S.D.) of detected camera signal versus mean number of camera signal for an intensity-stabilized laser impinging on the EMCCD camera.
        Fitting the expected square root scaling provides the conversion factor between camera signal to photon number to be $0.32(3)$.
    }
    \label{fig:s9}
\end{figure}
Deriving the conversion of the camera signal (counts) to incident photons, $\alpha = N_p/c$ with $N_p$ being the photon number and $c$ being the camera signal (counts), requires knowledge of the camera performance.
We performed two independent measurements to estimate the EMCCD conversion factor $\alpha$ quantifying the conversion between detected photons and recorded counts.
For the first method, we illuminated the EMCCD camera with an intensity-stabilized laser beam and compared the camera signal for settings with and without camera gain.
Here, we retrieved the camera signal to photon conversion factor of $\alpha = 0.298(1)$, including the camera quantum efficiency of $0.8$ and sensor conversions.
The second method exploits the scaling of the photon shot noise, $\Delta N_p = \sqrt{N_p}$.
As a consequence, the relation of the standard deviation of the camera signal and the camera signal is $\Delta c = \sqrt{2/\alpha}\cdot \sqrt{c}$.
Here, the factor of $\sqrt{2}$ is the additional noise originating from the amplification process of the EMCCD camera~\cite{Robbins2003}.
Varying the light intensity, we extracted the conversion factor of $\alpha = 0.32(3)$ from a fit, see Fig.~\ref{fig:s9}.
Both are in agreement with each other; therefore, we use their average for analysing our data.

\end{document}